\documentclass[final]{svjour2}
\usepackage{graphicx}
\usepackage{rotating}
\usepackage{amssymb}
\usepackage{mathptmx}
\usepackage[numbers]{natbib}
\newcommand{\rhos}{\rho_{s}}
\newcommand{\rhon}{\rho_{n}}
\newcommand{\rhoo}{\rho_o}
\newcommand{\rhop}{\rho_{p}}
\newcommand{\mun}{\mu_{n}}
\newcommand{\rp}{r_{{p}}}
\newcommand{\up}{u_{{p}}}
\newcommand{\vn}{v_{{n}}}
\newcommand{\vs}{v_{{s}}}
\newcommand{\vvn}{\mathbf{v}_{n}}
\newcommand{\vvs}{\mathbf{v}_{s}}
\newcommand{\uup}{\mathbf{u}_{p}}
\newcommand{\rrp}{\mathbf{r}_{p}}
\newcommand{\VVs}{\mathbf{V}^\textrm{s}}
\newcommand{\VVb}{\mathbf{V}^\textrm{b}}
\newcommand{\nn}{\mathbf{n}}
\newcommand{\uu}{\mathbf{u}}
\newcommand{\vv}{\mathbf{v}}

\newcommand{\FF}{\mathbf{F}}
\newcommand{\VV}{\mathbf{V}}

\newcommand{\XX}{\mathbf{X}}

\newcommand{\ap}{a_{p}}

\newcommand{\rr}{\mathbf{r}}
\newcommand{\xx}{\mathbf{x}}

\newcommand{\bom}{{\mbox{\boldmath $\omega$}}}

\makeatletter
\journalname{Journal of Low Temperature Physics}
\bibpunct{}{}{,}{s}{}{,}

\begin{document}

\newcommand{\hdblarrow}{H\makebox[0.9ex][l]{$\downdownarrows$}-}
\title{Particles-vortex interactions and flow visualization in $^4$He}

\author{Y.A. Sergeev$^1$ and C.F. Barenghi$^2$}

\institute{$^1$School of Mechanical and Systems Engineering\\
$^2$School of Mathematics and Statistics\\ Newcastle University\\Newcastle upon Tyne,
NE1 7RU, England, United Kingdom\\ Tel.: +44 (0)191 2226284\\ Fax: +44 (0)191 2228600 \\
\email{yuri.sergeev@ncl.ac.uk}}

\date{XX.XX.2009}

\maketitle


\begin{abstract}

Recent experiments have demonstrated a remarkable progress in implementing and use of
the Particle Image Velocimetry (PIV) and particle tracking techniques for the study
of turbulence in $^4$He. However,
an interpretation of the experimental data in the superfluid phase requires
understanding how the motion of tracer particles is affected by the two components,
the viscous normal fluid and the inviscid superfluid. Of a particular importance is
the problem of particle interactions with
quantized vortex lines which may not only strongly affect the particle motion, but,
under certain conditions, may even trap particles on quantized vortex cores. The
article reviews recent theoretical, numerical, and experimental
results in this rapidly developing area of research, putting
critically together
recent results, and solving apparent inconsistencies.
Also discussed is a closely related technique of detection of
quantized vortices negative ion bubbles in $^4$He.

\end{abstract}

\section{Introduction and plan of the review} \label{introduction}

In this review we will be mostly 
concerned with the motion of small
solid particles in turbulent $^4$He. This new and rapidly
developing area of research has been initiated by the recent
success of implementation of the Particle Image Velocimetry (PIV) and the particle
tracking
techniques in superfluid helium, see Donnelly \emph{et
al.}\cite{D02}, Van Sciver and
co-workers\cite{C02,Z02,Z04,Z05_1,Z05_2}, and publications of the group involving
Paoletti, Bewley, Lathrop, Sreenivasan and their
co-workers\cite{B06,B08_a,B09_b,P08_a,P08_b}. (The only difference between these
two, otherwise identical, techniques is that the results of PIV yield the local
average particle velocities obtained by calculating cross-correlations of particle
ensembles and, therefore, result in the fluid-like, smooth velocity field, while the
particle tracking aims at investigation of individual particle trajectories.) These
are perhaps the only two well developed techniques which can
identify the flow patterns in turbulent superfluid helium. The PIV has been a standard
technique in classical fluid dynamics for several decades (see, for example, the
book by Raffel
\emph{et al.}\cite{R98}). In experimental studies of classical
turbulence, one can be confident that the motion of sufficiently
small particles will reveal the details of turbulent motion of the
viscous fluid. In superfluids, the two-fluid nature of $^4$He
makes an interpretation of PIV measurements much more difficult: A
solid particle interacts with both the normal fluid and the
superfluid; moreover, the particle may interact strongly with
quantized vortex lines in the superfluid component, and even become
trapped on them.

The mechanism of trapping of solid particles is essentially 
the same as that of trapping the charge carriers on quantized vortices
in $^4$He, primarily because e.g. the negative ion (electron) forms
around itself an almost macroscopic bubble of radius 
$12-20\,\textrm{\AA}$ from which helium atoms are excluded. The ion
trapping technique was used for detection of quantized vortices since
late 50s, and, although not being suitable for studying the normal
fluid patterns in $^4$He, can be considered as the technique closely
related to the PIV and the particle tracking methods. We will review
the development of the ion trapping technique in Section 7 of
this article.

The aims of experimental physicists working in the area of
superfluid turbulence are to understand the vortex tangle (quantum
turbulence), to measure velocity fluctuations in both the
normal fluid and the superfluid, and possibly to make
comparisons between quantum turbulence and ordinary turbulence.
Until recently the major
difficulty was lack of direct flow visualization technique near
absolute zero. Existing methods, such as measurements of
temperature differences to detect extra dissipation, ion trapping,
measurements of second sound attenuation by quantized vortices,
etc. probed only the vortex line density $L$ (total vortex length
per unit volume) averaged over an
experimental cell and did not reveal turbulent velocity
fluctuations (although a remarkable resolution has been recently
achieved in second sound measurements by Roche \emph{et
al.}\cite{Roche07} of the local tangle density).

Experimental breakthrough was made in 2002 when the PIV technique was successfully
introduced in
$^4$He (see references in the first paragraph of this Section).
This technique, which has been standard in classical fluid
dynamics for many years, is based on injecting many small
particles into the liquid. Two images are produced using short
laser pulses of different frequencies (corresponding e.g. to the green and red colors) focused into a narrow sheet and
separated in time by a few milliseconds. Software then analyzes
the images and identifies green and red dots corresponding to the
same particle at the two different times.
In this way the observed distance between the corresponding dots
yields the component of the local velocity in the plane of the
light sheet.

In classical fluids, provided the particle size is sufficiently
small (in turbulence studies, ``small'' means smaller than the
Kolmogorov length), the dominating force acting on the particle is
the viscous drag force, so that a researcher can be confident that
small particles trace the fluid motion (in particular, turbulent
velocity fluctuations) rather well.

Because the viscosity of liquid helium is very low the Kolmogorov
scale in the turbulent normal fluid can be very small, so that it
is essential that the particles used in visualization experiments are as
small as possible. In the cited experimental works the typical
particle diameter was of the order of $1\,\mu\textrm{m}$.

What do tracer particles trace in superfluid helium? One may
expect that, although the viscosity of $^4$He is low, the
dominating force exerted on the particle by the fluid will still
be the viscous drag, so that solid particles should trace the
normal fluid. However, this is not always true: due to the
two-fluid nature of superfluid helium, the particles interact not only
with the normal fluid, but also with the superfluid component
through inertial and added mass forces; moreover, the particles
interact strongly with quantized vortices in superfluid and may even be trapped on
superfluid vortex lines.
Therefore, if we want to interpret results of PIV and particle tracking measurements
correctly, we must answer first the question asked in the
beginning of this paragraph.

This article is divided in sections where theoretical and
numerical models of increasing complexity are compared to each other
and to experimental results. The first model, described in
Section 2, is the one-way coupling model. In Sections 2.1 and 2.2 we
derive the governing equations of motion of particle tracers in the presence
of two imposed fluids, the viscous normal fluid and the inviscid superfluid,
under the assumptions
that the particles do not disturb the flow, are smaller than any flow scale
of interest, and do not become trapped in vortices.
The one-way coupling model allows us to discuss the problem of
the stability of particles' trajectories, which is relevant
to the visualization a pure superflow. The one-way coupling model
is powerful enough to derive the general principles which
lead to
particles being trapped on vortices. In Section 2 we present
the experimental evidence for this trapping (Section 2.3),
and show how the mutual friction affects the motion of
particles near vortex cores (Section 2.4). Three case studies are
discussed of particle trajectories near vortices: vortex ring, thermal
counterflow tangles, and vortex tangles at low temperatures (Section 2.5).

Section 3 introduces the more sophisticated (and computationally more
expensive) two-way coupling model; in this model
the back reaction of the flow on the particle is taken into
account, and the dynamics of the particle-vortex interaction
and the trapping can be studied in great detail, including what happens
at the surface of the particle (which we assume to be spherical for
simplicity). Section 3.1 contains the
mathematical formulation of the two-way coupling model; Sections 3.2 and 3.3
are devoted to the numerical calculations of typical vortex-particle
interactions.

Section 4 makes use of the results of the numerical simulations described
in Section 3 to derive a simpler analytical model which explains
in a quantitive
way experiments performed in Florida\cite{C02,Z02,Z04,Z05_1,Z05_2}
 in which heavy particles fell through
a tangle of vortices. The two-way coupling
model is also applied to particles moving in tangles generated by a thermal
counterflow: these numerical results are applied to particle-tracking
experiments performed in Maryland\cite{B06,B08_a,B09_b,P08_a,P08_b}.
By considering numerical calculations
at small and high values of the vortex line density, we solve the apparent
disagreement between the Florida and Maryland experiments: the discussion
will reveal that the two 
experiments refer to two different regimes, which are both explained by the
two-way coupling model. In Section 4 we also discuss the experimental
observation of flow structures observed
 behind and in front a cylindrical obstacle
set in the middle of a counterflow channel, and propose a simple
analytical two-dimensional model which accounts, at least in principle, for the
qualitative existence of these flow structures.

Section 5 describes the most recent experimental
discoveries obtained using tracer particles:
the observation of turbulent boundary layer
flows, the visualization of individual vortex reconnections,
and the measurement of velocity statistics.
Section 7 reviews other techniques based on trapping ions and imaging
$\textrm{He}_2$ molecules; these techniques share important principles
(but not the size of the trapped object) with PIV and
particle tracking techniques.
Section 8 contains the final discussion.

\section{One-way coupling model of particle motion in turbulent
$^4$He} \label{one-way}

We begin answering this question with a relatively simple,
``one-way coupling'' model which follows the approach standard in
classical two-phase turbulence studies. We will generalize
to the two-fluid model of superfluid helium the equations of
motion of a solid spherical particle of radius $\ap$ in a
nonuniform flow of classical fluid. We assume that 1)~the presence
of particles does not modify the turbulence, 2)~the flow
velocities vary little over a distance of the order of particle
size, and 3)~particles do not interact strongly with quantized
vortex lines and are certainly not trapped on these lines. These
assumptions require that the particle size be much smaller than
both the Kolmogorov lengthscale, $b_\eta$ in the normal fluid, and
the mean intervortex distance, $\ell=L^{-1/2}$ in the superfluid.
Below we
formulate, under these assumptions, the equations of particle
motion.

\subsection{Fluid-particle interaction} \label{Fluid-particle_interaction}

We start with the fluid-particle interaction in classical liquids.
We consider a spherical solid particle of radius $\ap$ in the
nonuniform flow. Let the ambient (that is, in the absence of the
particle) fluid velocity field be $\vv(\rr,\,t)$. According to the assumptions
formulated above, the size of the particle is much smaller
than the flow lengthscale, $\lambda$, i.e. $\ap\ll \lambda$,
so that we can
introduce a small parameter
\begin{equation}
\varepsilon=\ap\|\nabla\vv\|/\vert\vv-\uup\vert\ll1\,, \label{varepsilon}
\end{equation}
where $\uup$ is the particle velocity.

\subsubsection{Fluid-particle interaction in the inviscid nonuniform flow}
\label{inviscid} As shown by Auton \emph{et al.}\cite{A88},
the total force acting on the particle in the nonuniform, inviscid
flow can, under assumption~(\ref{varepsilon}), be represented in the form
\begin{equation}
\FF=\FF^{(\textrm{i})} +\FF^{(\textrm{a})}+\FF^{(\omega)}\,,
\label{Inert_And_Added}
\end{equation}
where
\begin{equation}
\FF^{(\textrm{i})}=\rho\vartheta\,\frac{D\vv}{Dt} \qquad
\textrm{and} \qquad \FF^{(\textrm{a})}=C\rho\vartheta\,
\biggl(\frac{D\vv}{Dt}-\frac{d\uup}{dt}\biggr) \label{F}
\end{equation}
are the inertial and the added mass force, respectively, $\vartheta=\frac{4}{3}\pi\ap^3$ is the particle volume,
\begin{equation}
\frac{D}{Dt}=\frac{\partial}{\partial t}+(\vv\cdot\nabla)\,,
\label{DDt}
\end{equation}
and $C$ is the added mass coefficient (for spherical particle
$C=\frac{1}{2}$);
\begin{equation}
\FF^{(\omega)}=\frac{1}{2}\,\rho\vartheta(\vv-\uup)\times\bom\,,
\label{FOmega}
\end{equation}
where $\bom=\nabla\times\vv$ is the vorticity, represents the lift
force arising due to stretching of vortex lines in the vicinity of
the sphere's surface.

\subsubsection{Fluid-particle interaction in the nonuniform viscous flow}
\label{viscous}
We will consider the motion of a solid particle in the viscous
fluid assuming that the particle Reynolds number is small:
\begin{equation}
\textrm{Re}_p=\ap\vert\vv-\uup\vert/\nu\ll1\,,
\label{Rep}
\end{equation}
where $\nu$ is the kinematic viscosity. Note that small particle
Reynolds numbers are typical of PIV and particle tracking experiments, both in
classical
fluids and in superfluid helium. We will consider the particle
motion under the assumptions formulated above (in particular that
the parameter $\varepsilon$ introduced by
formula~(\ref{varepsilon}) is small).

Detailed analysis of the forces acting on the particle in the
nonuniform viscous flow can be found in works by Maxey and
Riley\cite{Ma83}, Mei\cite{Me94}, and Kim, Elghobashi, and Sirignano\cite{Ki98}. The total force acting on the particle can be
represented as a sum of several contributions, i.e. the gravity, viscous drag, the
inertial and added mass force, Fax\'en correction arising due to the local
non-uniformity of the ambient flow, the Saffman lift force arising due to the local
shear, and the Magnus lift force arising due to rotation of the particle.

The main contribution, dominating in most particulate flows, is
the viscous Stokes drag force:
\begin{equation}
\FF^{(\textrm{d})}=6\pi\ap\rho\nu(\vv-\uup)\,. \label{Stokes}
\end{equation}

Surprisingly, in the viscous flow the inertial and added mass
forces, $\FF^{(\textrm{i})}$ and $\FF^{(\textrm{a})}$ are
determined by the same formulae~(\ref{Inert_And_Added})-(\ref{F})
as for the inviscid flow, with the same added mass coefficient,
$C=\frac{1}{2}$ for the spherical particle.

\subsubsection{Fluid-particle interaction in $^4$He} \label{force_He}
To determine the force acting on the particle in $^4$He, we
simply add together all the forces exerted by the normal fluid and
the superfluid. Since the superflow is potential, $\FF^{(\omega)}_s$ is identically
zero, provided the particle does not become trapped on
quantized vortex lines. As shown by Poole \emph{et al.}\cite{P05}, for
the flow properties and physical parameters typical of the normal component of
$^4$He, the Fax\'en correction, the history and lift forces can be neglected
provided the particle Reynolds number and the parameter $\epsilon$ are small and the
particle size is significantly smaller than the Kolmogorov length. Therefore, the
total force acting on the particle can be approximated as
\begin{equation}
\FF=\FF^{(\textrm{g})}
+\FF^{(\textrm{d})}_n+\FF^{(\textrm{i})}_n
+\FF^{(\textrm{a})}_n+\FF^{(\textrm{i})}_s+\FF^{(\textrm{a})}_s\,,
 \label{force}
\end{equation}
where the subscripts $n$ and $s$ refer to the normal fluid and
the superfluid, respectively. Here $\FF^{(\textrm{g})}$ is the
combination of gravity and buoyancy, and $\FF^{(\textrm{d})}_n$
is the viscous drag force exerted by the normal fluid:
\begin{equation}
\FF^{(\textrm{g})}=\vartheta(\rhop-\rho)\mathbf{g}, \qquad
\FF^{(\textrm{d})}_n=6\pi\ap\mun(\vvn-\uup)\,,
\label{gravityANDstokes}
\end{equation}
where $\mun$ is the viscosity of $^4$He.
Substantial derivatives required for determining the inertial and
added mass forces in the normal and superfluid components are now defined by
formula~(\ref{DDt}) with $\vv$ replaced by $\vvn$ and $\vvs$, respectively.

\subsection{Lagrangian equations of particle motion}
\label{Lagrangian_equations}

We arrive at the following equation of particle motion\cite{P05}:
\begin{eqnarray}
&&\rhop\vartheta\,\frac{d\uup}{dt}=6\pi\ap\mun(\vvn-\uup)+\vartheta(\rhop-\rho)\mathbf{g}
\nonumber\\ &&+\rhon\vartheta\,\frac{D\vvn}{Dt}
+C\rhon\vartheta\,\biggl(\frac{D\vvn}{Dt}-\frac{d\uup}{dt}\biggr)
+\rhos\vartheta\,\frac{D\vvs}{Dt}
+C\rhos\vartheta\,\biggl(\frac{D\vvs}{Dt}-\frac{d\uup}{dt}\biggr)\,.
\label{motion1}
\end{eqnarray}
This equation must be considered together with the kinematic
equation
\begin{equation}
d\rr/dt=\uup\,, \label{kinematic}
\end{equation}
where $\rr=\rr(t)=(x(t),\,y(t),\,z(t))$ ($\equiv\rrp(t)$)
should be regarded as a Lagrangian trajectory of the solid
particle. Eqs.~(\ref{motion1}) and (\ref{kinematic})
constitute the closed system for the unknown particle position and
velocity, $\rr(t)$ and $\uup(t)$, respectively.

Often (although not always) in the PIV and particle tracking
experiments neutrally buoyant particles (with $\rhop=\rho$) are used in
order to eliminate unwanted effects of gravity on the particle
motion. For neutrally buoyant particles, Eq.~(\ref{motion1})
can be written in a more concise form
\begin{equation}
\frac{d\uup}{dt}=\frac{1}{\tau}(\vvn-\uup)
+\frac{3}{2\rhoo}\biggl(\rhon
\frac{D\vvn}{Dt}+\rhos\frac{D\vvs}{Dt}\biggr)\,, \label{motion2}
\end{equation}
where
\begin{equation}
\rhoo=\rhop+\rho/2=3\rho/2 \quad \textrm{and}
\quad \tau=2\ap^2\rhoo/(9\mun)\,. \label{rho_tau}
\end{equation}
The parameter $\tau$, which shows how quickly the particle adjusts its motion to the
viscous flow, plays an important r\^ole in the study of
particle motion in fluids, and is commonly known as either the
particle response time, or viscous relaxation time.

Analyzing Eqs.~(\ref{kinematic}) and (\ref{motion2}), Poole \emph{et al.}\cite{P05} showed
that, provided $\tau/\tau_f\ll1$, where
$\tau_f$ is the timescale of the fluid motion (e.g. the
Kolmogorov time), the neutrally buoyant particle tracks the motion of the normal
fluid. If $\tau/\tau_f\gg1$, the particle moves with a velocity corresponding to the
total current density, $\textbf{j}=\rhon\vvn+\rhos\vvs$.

However, two important issues may invalidate these conclusions: 1) instability of
particle
trajectories, and 2) trapping of particles on superfluid vortex
lines. The analysis of particle trapping will require more
elaborate, self-consistent, two-way coupling model which would
account for deformation of the vortex filament by the approaching
particle, including possible reconnection of the vortex with the
particle surface (see below Section~\ref{self-consistent}).

\subsubsection{Instability of particle trajectories} \label{instability}

Instability of Lagrangian trajectories of the neutrally buoyant
particle in the classical viscous fluid was discovered and studied
relatively recently by Babiano \emph{et al.}\cite{Bab00}. To
illustrate such an instability in turbulent superfluid at finite
temperature such that both the normal fluid and the superfluid
component are present, we consider the simplest case assuming that
$\vvn\approx\vvs=:\vv$ down to the length scales comparable to the
vortex line spacing. The Lagrangian equations of particle motion,
which become
\begin{equation}
\frac{d\uup}{dt}=\frac{\vv-\uup}{\tau} +\frac{D\vv}{Dt}\,, \qquad
\frac{d\rrp}{dt}=\uup\,, \label{motion-turbo}
\end{equation}
have, provided $\uup(0)=\vv(0)$ and $\rrp(0)=\rr_f(0)$, a
formal solution $\uup(t)=\vv(\rr_f,\,t)$,
$\rrp(t)=\rr_f(t)$, where $\rr_f(t)$ is a
trajectory of the fluid point. Therefore, it would seem natural to
conclude that neutrally buoyant particles follow the fluid
exactly.

\begin{figure}
\begin{center}
\includegraphics[%
  width=0.75\linewidth,
  keepaspectratio]{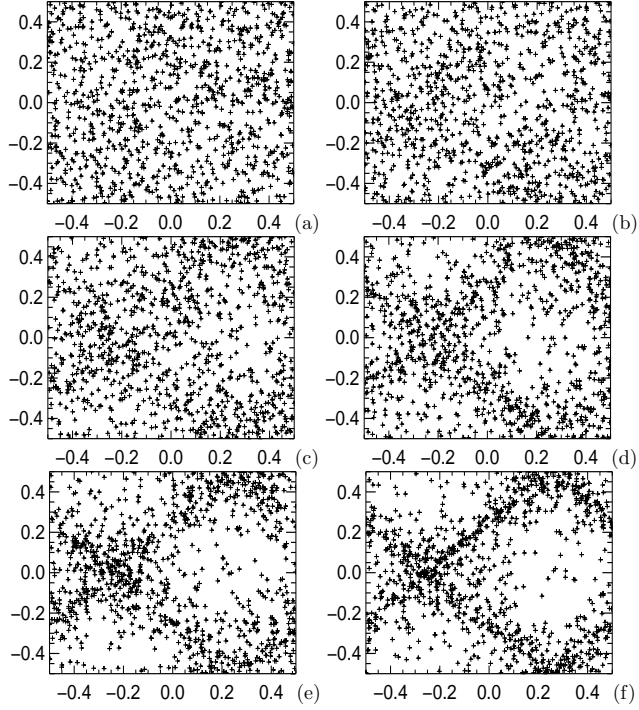}
\end{center}
\caption{Positions of tracer particles at different times\cite{P05}.
From Poole, Barenghi, Sergeev, and Vinen,
\emph{Phys. Rev. B}, \textbf{71}, 064514, (2005). Reprinted by permission, \copyright2005 American Physical
Society.} \label{segregation}
\end{figure}

However, let us consider now the particle motion in the so-called
ABC (Arnold-Beltrami-Childress) flow, frequently used as the
simplest model of turbulence, whose velocity field is
\begin{eqnarray}
&&v_x=A \sin(2 \pi z)+C \cos(2 \pi y)\,,\quad v_y=B \sin(2
\pi x)+A \cos(2 \pi z)\,,\nonumber \\ &&v_z=C \sin(2 \pi
y)+B \cos(2 \pi x)\,. \label{ABC_eqs}
\end{eqnarray}
The time sequence\cite{P05} illustrating the position
of large number of neutrally buoyant tracer particles in the ABC flow, starting
from their uniform distribution, is shown in
Fig.~\ref{segregation}. It can be seen that particles do not
follow the fluid, but instead, due to instabilities of their
trajectories, travel from the regions of high vorticity to the
regions of high rate of strain. (In classical multiphase fluid
dynamics similar phenomenon for particles heavier than the fluid
has been studied in detail. However, the mechanism of segregation
of heavy particles is quite different from that of neutrally
buoyant particles.) Note, though, that this instability develops
rather slowly: the last frame of Fig.~\ref{segregation}, recovering the lines of
minimum vorticity and maximum rate of strain,
corresponds to 3 times of turnover of the ABC flow.

\begin{figure}
\begin{center}
\includegraphics[%
  width=0.45\linewidth,
  keepaspectratio]{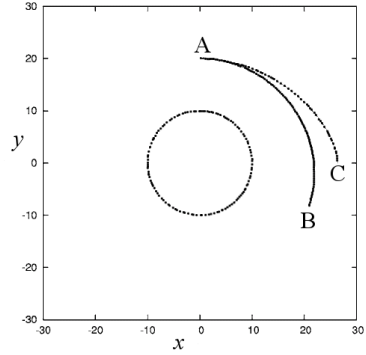}
\end{center}
\caption{Trajectories\cite{S06_a} of neutrally buoyant solid particle
(dashed line A to C) and superfluid particle (solid line A to B)
around three vortices moving along the closed orbit shown.
From Sergeev, Barenghi, Kivotides, and Vinen,
\emph{Phys. Rev. B}, \textbf{73}, 052502, (2006).
Reprinted by permission, \copyright2006 American
Physical Society.} \label{three_vortices}
\end{figure}

Instability of particle trajectories is more pronounced in the
case $T\to0$ when the normal fluid is absent. The equations of
motion of neutrally buoyant particle become
\begin{equation}
\frac{d\uup}{dt}= \frac{\partial \vvs}{\partial t}
                         +(\vvs \cdot \nabla) \vvs=-\nabla p\,,
                         \quad \frac{d\rrp}{dt}=\uup\,,
\label{motionTto0}
\end{equation}
and have a formal solution $\uup(t)=\vvs(\rr_s(t),\,t)$,
$\rrp(t)=\rr_s(t)$, where
$d\rrp/dt=\vvs(\rr_s,\,t)$, so that, in the case of very
low temperature when the normal fluid is absent, it can be
expected that neutrally buoyant particles trace the superfluid.
However, this is not the case either. Consider the
simplest case of the neutrally buoyant particle moving around a
single, stationary, straight vortex line. In cylindrical polar
coordinates $(\rp,\,\theta_p)$ the equations of particle
motion are:
\begin{equation}
\ddot{r}_p-\rp\omega_p^2=-\kappa^2/(2\pi^2\rp^3)\,,
\qquad 2\omega\dot{r}_p+\rp\dot\omega_p=0\,,
\label{polar}
\end{equation}
where $\omega_p=\dot{\theta}_p$. If at the
initial moment the particle velocity does not coincide exactly
with the fluid velocity, the particle will spiral either outwards, or inwards. We
arrive at the conclusion which remains valid in the general case
of particle motion at temperature $T\to0$ (Sergeev \emph{et
al.}\cite{S06_a}): unless the initial velocity of neutrally
buoyant particle matches exactly the velocity of the fluid point,
the trajectory of the solid particle deviates significantly from
the trajectory of the fluid point. Such an instability is
amplified by any macroscopic mismatch between the velocity of the
superfluid and the velocity of the particle at the beginning of
the experiment. This is further illustrated by
Fig.~\ref{three_vortices} showing the trajectories of solid and
fluid particles around three vortices.

\subsection{Trapping of particles on quantized vortices:
mechanism and experimental evidence} \label{trapping}

\subsubsection{Mechanism} \label{trapping_mechanism}

Why can the superfluid vortex trap the particle? To answer this
question, we have to take into account a possibility of
reconnection of the quantized vortex to the surface of moving
particle (a more detailed analysis of the mechanism of this
process will be given below in Section~\ref{self-consistent}).

Below three different versions, or, rather, three different ways
of explaining the reason for particle trapping are suggested.

\begin{figure}
\begin{center}
\includegraphics[%
  width=0.55\linewidth,
  keepaspectratio]{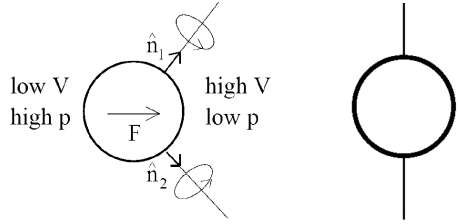}
\end{center}
\caption{Asymmetric reconnection of the vortex to the particle
surface creates a force restoring the symmetric particle-vortex
configuration\cite{S06_b}. From Sergeev, Barenghi, and Kivotides, \emph{Phys. Rev.
B}, \textbf{74}, 184506, (2006). Reprinted by permission,
\copyright2006 American Physical Society.}
\label{asymmetric}
\end{figure}

\noindent{$1^o$.} Imagine that the quantized vortex reconnects
symmetrically to the surface of spherical particle, as shown in
the right part of Fig.~\ref{asymmetric}. The kinetic energy of the flow
field created by the straight quantized vortex can be easily
calculated, and in the symmetric configuration
is reduced by
\begin{equation}
\Delta E=\frac{\rhos\kappa^2\ap}{2\pi}\,\ln\frac{2\ap}{\xi}\,,
\label{reduced_E}
\end{equation}
where $\xi\approx10^{-8}\,{\rm cm}$ is the vortex core radius.
Note that assuming $\ap\gg\xi$ this result follows from the
substitution energy calculated by Parks and
Donnelly\cite{Parks-Donnelly} for the ion bubble trapped by the
quantized vortex line, see below Sec.~\ref{related_techniques} and
formula~(\ref{substitution}) therein. Formula~(\ref{reduced_E})
determines the kinetic energy which the particle would require to
break free from the symmetric vortex configuration shown in
Fig.~\ref{asymmetric} (right).

\noindent{$2^o$.} The flow field of the vortex creates a pressure
gradient
\begin{equation}
\nabla
p=-\rhos(\vvs\cdot\nabla)\vvs=\frac{\rhos\kappa^2}{8\pi^2}\,\nabla\biggl(\frac{1}{r^2}\biggr)
\label{nabla_p}
\end{equation}
attracting the particle to the vortex.

\noindent{$3^o$.} If the particle-vortex configuration is
symmetric, as in the right part of Fig.~\ref{asymmetric}, then,
obviously, the force acting on the particle is zero. Now imagine
that this symmetry is perturbed as shown in the left part of this
figure. The superfluid component is inviscid, so that the
following arguments apply based on Bernoulli's integral: the fluid
velocity on the right side of the particle surface, where the two
vortex strands are closer, is greater than that on the left side,
and, therefore, the pressure is bigger on the left side of the
sphere. This provides a net force restoring the symmetric
particle-vortex configuration.

\subsubsection{Evidence of particle trapping} \label{evidence}

Experimental evidence comes from the publication
of Bewley, Lathrop and Sreenivasan\cite{B06}
(see also more recent papers\cite{B09_b,P08_a,P08_b}).
\begin{figure}
\begin{center}
\includegraphics[%
  width=0.75\linewidth,
  keepaspectratio]{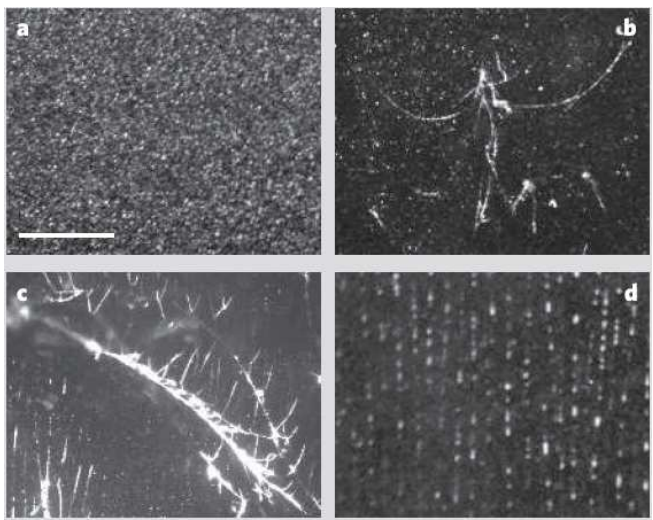}
\end{center}
\caption{PIV visualization\cite{B06} by Bewley, Lathrop, and Sreenivasan
(\emph{Nature},
\textbf{441}, 588, (2006)) of
quantized vortex cores: (a) -- above $\lambda$ transition; (b) and
(c) -- branching filaments tenth of mK below $\lambda$ transition;
(d) -- regrouping along vertical lines in steady rotating
apparatus. Reprinted by permission, \copyright2006 Macmillan Publishers Ltd.}
\label{Sreeni_Nature}
\end{figure}
Fig.~\ref{Sreeni_Nature}, published in the cited paper\cite{B06}, shows that the
researchers, using
the PIV technique, actually ``painted'' quantized vortices by
tracer particles.

\begin{figure}
\begin{center}
\includegraphics[%
  width=0.5\linewidth,
  keepaspectratio]{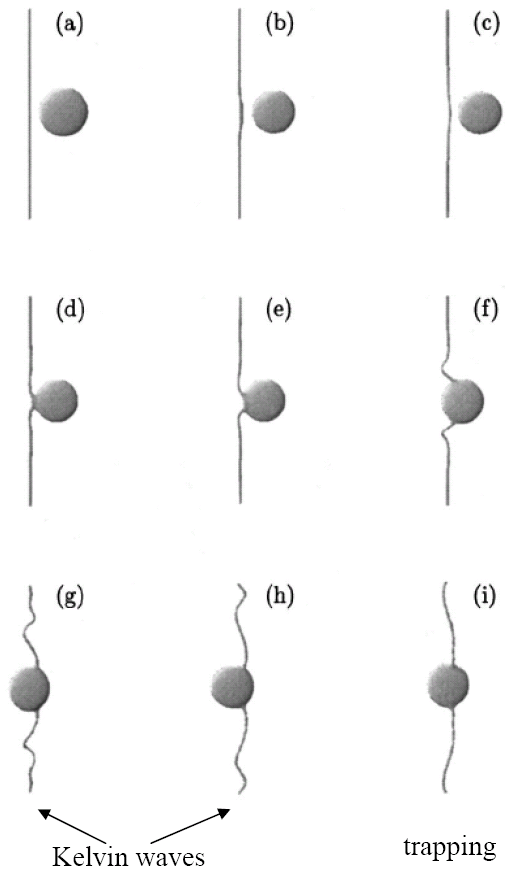}
\end{center}
\caption{Trapping of the ion bubble by the quantized vortex:
microscopic calculation\cite{Be00} by Berloff and Roberts
(\emph{Phys. Rev. B}, \textbf{63},
024510, (2000)) based on the
Gross-Pitaevskii equation. Reprinted by permission, \copyright2000 American
Physical Society.} \label{ion}
\end{figure}

Another evidence comes not from experiments, but from the quantum
calculation based on the Gross-Pitaevskii equation. Berloff and
Roberts\cite{Be00} studied an interaction between the negative ion
and the quantized vortex. In superfluid helium, the negative ion
forms around itself a bubble of diameter approximately
$16\times10^{-8}\,\textrm{cm}$ (an order of magnitude higher than
the size of the vortex core, $\xi\approx10^{-8}\,\textrm{cm}$)
and, therefore, can be treated, for our purpose, as a particle
(albeit very small). Results of Berloff and Roberts' calculation
are illustrated in Fig.~\ref{ion}: the ion bubble approaches the
vortex, which deforms ((b) and (c)) and then reconnects to the
particle surface (d); the reconnection excites Kelvin waves ((f),
(g), and (h)) which carry away the energy, and, eventually, the
bubble-vortex configuration relaxes and the ion becomes trapped on
the vortex core (i).

\subsection{Mutual friction and trapping} \label{mutual_friction}

At this point, the question can be asked whether it is worth or
not further exploiting the one-way coupling model which
neglects any influence of the particle on the vortex evolution.
The answer is ``yes'': there are cases
where trapping events are not very frequent, so that a useful information
about the behaviour of tracer particles can be obtained by ignoring their
trapping on vortex cores. We will also show that the mutual friction between
quantized vortices and the normal fluid can prevent trapping.

\begin{figure}
\begin{center}
\includegraphics[%
  width=0.75\linewidth,
  keepaspectratio]{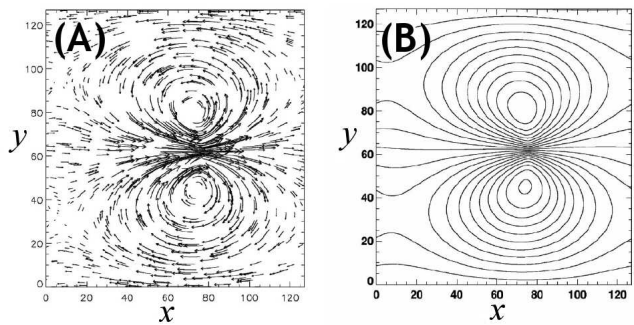}
\end{center}
\caption{Velocity field of the normal fluid due to the mutual
friction forcing of a single superfluid vortex\cite{I00_b}. (A):
velocity arrow plot; (B): streamlines. From
Idowu, Willis, Barenghi, and Samuels,
\emph{Phys. Rev. B}, \textbf{62}, 3409, (2000).
By permission, \copyright2000
American Physical Society.}
\label{normal_disturbances}
\end{figure}

\begin{figure}
\begin{center}
\includegraphics[%
  width=0.45\linewidth,
  keepaspectratio]{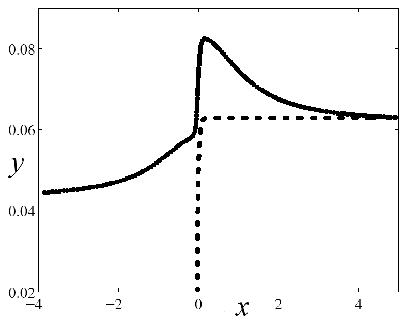}
\end{center}
\caption{Particle trajectories\cite{S07} in the presence (solid lines) and
absence (dashed line) of the normal fluid disturbances.
From Sergeev, Wang, Meneguz, and Barenghi, \emph{J. Low
Temp. Phys.}, \textbf{146}, 417, (2007).
Reprinted by permission, \copyright2007 Springer.}
\label{deflection}
\end{figure}

In the vicinity of the vortex core, the mutual friction induces,
in the normal fluid, the vortex dipole whose typical lengthscale
is expected to be about $100\,\mu\textrm{m}$. The results of
numerical calculation by Idowu \emph{et al.}\cite{I00_b} of the dipole-like normal
fluid disturbance are shown in Fig.~\ref{normal_disturbances}.
This normal flow disturbance can deflect the tracer particle which
otherwise would have collided with, and possibly trapped by the
vortex. Typical trajectory, calculated by Sergeev \emph{et
al.}\cite{S07}, of the particle
moving from right to left and interacting with the superfluid
vortex and normal fluid disturbance is shown in
Fig.~\ref{deflection} by the solid line. For
comparison, the trajectory calculated without taking into account
the normal fluid disturbance is shown by the dashed line. This
trajectory leads to the collision with the vortex core located at
the origin.

\begin{figure}
\begin{center}
\includegraphics[%
  width=0.25\linewidth,
  keepaspectratio]{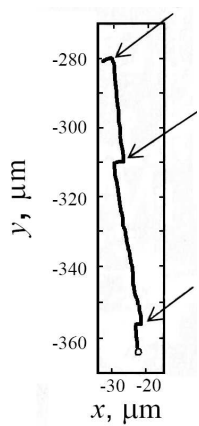}
\end{center}
\caption{Observation by Bewley, Lathrop, Sreenivasan, and Paoletti\cite{B07} of
particle trajectories
possibly indicating an influence of normal fluid disturbances
induced by the mutual friction.} \label{jostling}
\end{figure}

Normal fluid vortical structures induced by the mutual
friction were predicted by Hall and Vinen\cite{H56}
and Kivotides, Barenghi, and Samuels\cite{Science2000},
but so far, because of low
resolution of experimental techniques, there was no direct
experimental proof of existence of the normal fluid disturbances.
Owing to the much higher resolution, the PIV and particle tracking techniques can
provide such an evidence. Perhaps the
first experimental confirmation of existence of normal flow
structures induced by the mutual friction has already been found.
Fig.~\ref{jostling} shows a typical particle trajectory observed
by Bewley, Lathrop, Sreenivasan, and Paoletti\cite{B07}, by means of
the particle tracking technique, in the thermal counterflow. The
particle trajectory has characteristic `kinks' resembling the
`deflected' trajectory shown by the solid line in
Fig.~\ref{deflection}.

In the remaining part of this Section we will analyze, neglecting
trapping, three examples of particle motion in $^4$He.

\subsection{Case studies} \label{case}

\subsubsection{Vortex ring propagating against a particulate
sheet} \label{ring_particulate_sheet}

\begin{figure}
\begin{center}
\includegraphics[%
  width=0.65\linewidth,
  keepaspectratio]{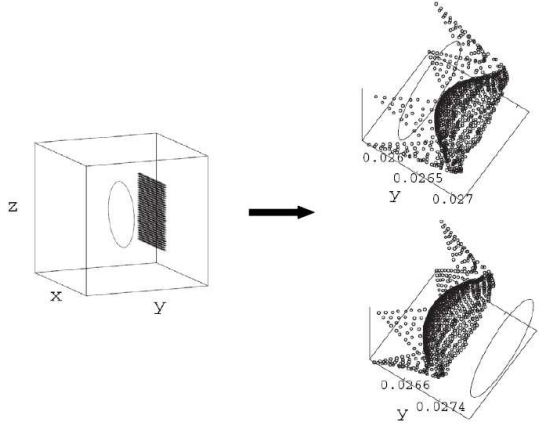}
\end{center}
\caption{Vortex ring and solid particles' configurations before and
after the ring has passed through the particulate sheet\cite{K05}.
Adapted from Kivotides, Barenghi, and Sergeev, \emph{Phys. Rev.
Lett.}, \textbf{95}, 215302, (2005). By
permission, \copyright2005 American Physical Society.} \label{ring_sheet}
\end{figure}

This study\cite{K05} led to a proposal of a simple experiment
(not performed yet) which could, in principle, justify the use of
PIV technique for measuring instantaneous normal fluid velocity
patterns. We consider a single vortex ring propagating normally to a
plane sheet of neutrally buoyant particles, see
Fig.~\ref{ring_sheet} (left).

In the considered model, the mutual friction between the normal
fluid and the superfluid vortex is taken into account leading to
emergence of the normal fluid disturbances described earlier. The
motion of the superfluid vortex whose core is defined parametrically
by $\XX(s,\,t)$ is governed by the following equation derived by
Idowu \emph{et al.}\cite{I00_a}:
\begin{equation}
\partial\XX/\partial
t=\vv_\ell=h\VVs+h_*\XX'\times(\vvn-\VVs)-h_{**}\XX'\times(\XX'\times\vvn)\,,
\label{vortex_motion_1}
\end{equation}
where $\XX'=\partial\XX/\partial s$, $h(T)$, $h_*(T)$ and
$h_{**}(T)$ are the known mutual friction coefficients, and the
vortex-induced superfluid velocity $\VVs$ is given by the
Biot-Savart integral
\begin{equation}
\VVs(\xx)=-\frac{\kappa}{4\pi}\int
ds\,\frac{\XX'\times(\XX-\xx)}{\vert\XX-\xx\vert^3}\,.
\label{Biot-Savart_1}
\end{equation}
The motion of incompressible ($\nabla\cdot\vvn=0$) normal fluid is
governed by the equation
\begin{equation}
\frac{\partial\vvn}{\partial
t}+(\vvn\cdot\nabla)\vvn=-\frac{1}{\rho}\nabla
p+\nu\nabla^2\vvn+\frac{1}{\rho}\FF\,, \label{N-S}
\end{equation}
where $\FF$ is the mutual friction force per unit volume. This
force is determined as the sum of the drag force and the
Iordanskii force, $\mathbf{f}$, per unit length:
\begin{equation}
\mathbf{f}=\rhos\kappa[d_{**}\XX'\times(\XX'\times(\vvn-\VVs))-d_*\XX'\times(\vvn-\VVs)]\,.
\label{Iordanskii}
\end{equation}
Here the new mutual friction coefficients, $d_*(T)$ and $d_{**}(T)$
are known and can be expressed explicitly through $h(T)$, $h_*(T)$
and $h_{**}(T)$.

\begin{figure}
\begin{center}
\includegraphics[%
  width=0.75\linewidth,
  keepaspectratio]{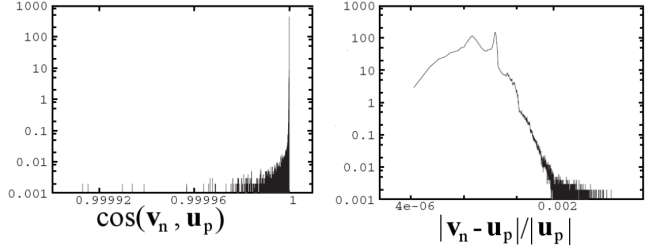}
\end{center}
\caption{Histograms\cite{K05} of the cosine of the angle
between $\vvn$ and $\uup$ (left) and of the magnitude of relative
velocity, $\vert\vvn-\uup\vert$ (right). The vertical axes are
divided by 1000. From Kivotides, Barenghi, and Sergeev, \emph{Phys. Rev.
Lett.}, \textbf{95}, 215302, (2005).
Reprinted by permission, \copyright2005 American Physical
Society.} \label{histograms_sheet}
\end{figure}

Trajectories of neutrally buoyant particles were found by numerical
integration of Eqs.~(\ref{kinematic}) and (\ref{motion2}).
Configurations of solid particles before and after the ring has
passed the particulate sheet are shown in Fig.~\ref{ring_sheet}
(right).
Fig.~\ref{histograms_sheet} shows the histograms of the angle
and of the relative difference between $\vvn$ and $\uup$; it can be
seen that $\vvn$ and $\uup$, to a very good degree of accuracy, are
identical both in magnitude and direction. Moreover, the calculation
showed that trapping events are relatively rare -- only 42 out of
900 particles approached the ring to a distance smaller than three
particle diameters. This enables us to conclude that, in the
proposed experiment, the measurement of particle velocities can
provide direct information about instantaneous normal flow patterns.

Later, based on the self-consistent, two-way coupling model, Kivotides and
Wilkin\cite{K_Wil08} performed more elaborate study of interactions between solid
particles and vortex rings, see below Sec.~\ref{Kivotides_Wilkin}.

\subsubsection{Particle motion in thermal counterflow} \label{counterflow_one_way}

In this example\cite{K06} we will be concerned with the T-I state
of $^4$He turbulence such that the vortex tangle in the
superfluid component is present but the normal flow is laminar.
For simplicity, the normal flow is assumed uniform,
$\vvn=\textbf{const}$. The superfluid velocity can be represented
as $\vvs=\vvs^C+\VVs$, where the mean (counterflow) superfluid
velocity, $\vvs^C$ is linked with the normal fluid velocity by the
relation $\rho\vvs^C+\rhon\vvn=\mathbf{0}$, and $\VVs$ is the
fluctuating superfluid velocity induced by the vortex tangle. The
dynamic vortex tangle is modeled, in the periodic box, taking into
account the mutual friction between the normal fluid and quantized
vortices, as well as the Biot-Savart interaction between vortex
filaments. An influence of superfluid vortices on the motion of
normal fluid is neglected. Based on Eqs.~(\ref{kinematic}) and
(\ref{motion2}), the motion of neutrally buoyant tracer particles
of diameter $d_p=6.25\times10^{-3}\,\textrm{cm}$ was
calculated after the vortex tangle has reached the statistically
steady state. Calculations, performed at temperature
$T=1.3\,\textrm{K}$ for $\vert\vvn\vert=1.1417$ and
$0.6058\,\textrm{cm}/\textrm{s}$ (corresponding values of the
counterflow heat flux are $q=1.07\times10^{-3}$ and
$4.57\times10^{-4}\,\textrm{J}/(\textrm{cm}^2\cdot\textrm{s})$),
and at temperature $T=2.171\,\textrm{K}$ for
$\vert\vvn\vert=0.01183\,\textrm{cm}/\textrm{s}$
($q=0.125\,\textrm{J}/(\textrm{cm}^2\cdot\textrm{s})$), showed
that the particle velocity is very narrowly peaked around the
constant normal velocity.

\begin{figure}
\begin{center}
\includegraphics[%
  width=0.95\linewidth,
  keepaspectratio]{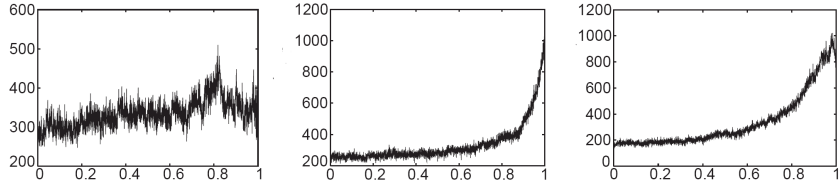}
\end{center}
\caption{Histograms\cite{K06} of $\vert\cos(\VVs,\,\uup)\vert$.
From Kivotides, Barenghi, and Sergeev, \emph{Europhys.
Lett.}, \textbf{73}, 733, (2006). Reprinted by
permission, \copyright2006 EDP Sciences.} \label{histograms_cos_counterflow}
\end{figure}

Three histograms of Fig.~\ref{histograms_cos_counterflow},
corresponding to the three cases considered above, show the absence
of alignment (left) between particle velocities and the superfluid
velocity, $\VVs$ induced by the vortex tangle, and 
illustrate anisotropy in $\VVs$ (center and right).

The above two examples enable us to expect that, in the range of
parameters typical of PIV measurements, neutrally buoyant
particles should trace the normal fluid well, and, provided strong
interactions of particles with quantized vortices can be
neglected, particle velocity fluctuations induced by interactions
between particles and quantized vortices should be relatively small.
However, these conclusion can be invalidated by trapping of solid
particles on quantized vortex cores. This phenomenon will be
addressed below in the Sec.~\ref{self-consistent}.

\subsubsection{Particle motion in a vortex tangle at very low
temperature} \label{Tto0}

Below the results discussed 
in Section~\ref{temperature} will suggest that at
temperature $T<0.5\,\textrm{K}$, when the normal fluid is
practically absent and the damping force on the particle can be
neglected, trapping of neutrally buoyant solid particles on
quantized vortices can, most likely, 
be ignored and the motion of solid particle
can be modeled by a simpler, one-way coupling model. However, at
these temperatures the presence and motion of the particles still
affects the vortex filaments, so that a certain modification
should be necessary of the one-way coupling model. In order to
understand some features of the particle motion in the vortex
tangle at such a low temperature, we start with a simple,
two-dimensional model of the tangle\cite{K08_PoF}. In such a model
the vortex lines become vortex points, and the Biot-Savart law
reduces to a simple statement that each vortex point moves as a
fluid point in a flow field created by all other vortices. (Such a
two-dimensional system of vortex points is known as the Onsager's
point vortex gas\cite{O49}.)

We consider, in the periodic box, the motion of neutrally buoyant
particles in the system of vortex points of random polarity. We
neglect trapping of particles on quantized vortices as well as any
influence of particles on the motion of vortex points.

In this approximation the particle motion is governed by Eqs.~(\ref{motionTto0}), and the following problem can be immediately
identified: in the case where the position of the particle
coincides with that of the vortex point, the pressure gradient
force, $-\nabla p$ becomes unphysically singular. Based on the
mechanism, described below in Sec.~\ref{self-consistent} of the
particle-vortex collision at very low temperature (in the absence
of damping force), we will resolve this manifest difficulty by
modifying the model~(\ref{motionTto0}) as follows.

The cause of the problem is that in the real,
three-dimensional tangle at temperature $T<0.5\,\textrm{K}$, even
when the particle breaks through the vortex, it nevertheless
reconnects with the vortex filament when the distance between the
particle and the vortex core becomes of the order of particle
radius. Since the vortex line attached to the particle is
necessarily orthogonal to the particle surface, the reconnection
results in a dramatic decrease of the force exerted on the
particle; this force is zero when the particle-vortex
configuration is symmetric. Then, the one-way coupling model can
be modified by assuming that there exists a force-free region for
$\rp<a_c$, where $a_c\sim O(\ap)$ is a cut-off distance.

\begin{figure}[t]
\begin{tabular}[b]{cc}
\includegraphics[height=0.32\linewidth]{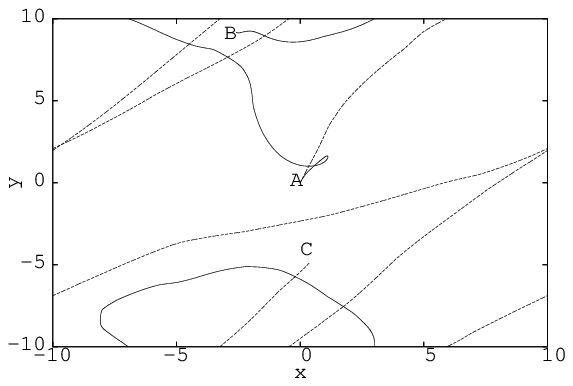}&
\includegraphics[height=0.31\linewidth]{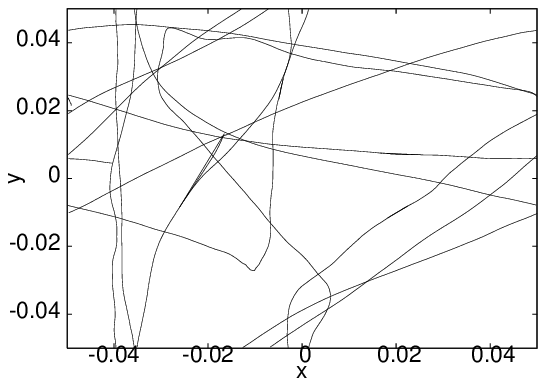}\\
\end{tabular}
\caption{Left: trajectories of a solid particle (dashed line) and of a
fluid point (solid line) in a system of 20 vortex points. Right: projection of the
particle trajectory in the three-dimensional vortex tangle\cite{K08_PoF}.
From Kivotides, Sergeev, and Barenghi, \emph{Phys. Fluids}, \textbf{20},
055105, (2008). Reprinted by
permission, \copyright2008 American Institute of Physics.}
\label{Tr2D_Tr3D}
\end{figure}

In the two-dimensional calculation we set $a_\textrm{c}=\ap$.
Trajectories of an inertial particle and a fluid point are
illustrated in Fig.~\ref{Tr2D_Tr3D} (left) for a system of 20
vortex points
set, initially, at random locations. The solid particle starts its
motion at the point A where it has the velocity equal that of the
fluid particle, but very quickly the trajectory of the solid
particle looses any resemblance to the trajectory of the fluid
point; moreover, the trajectory of the solid particle soon
acquires the ballistic character. The reason for such a behaviour
of the solid particle is the instability of its trajectory as
discussed in Sec.~\ref{instability}.

The evolution of particle velocity with time reveals another,
rather unexpected feature: although
$\vert\uup(0)\vert=\vert\vvs(0)\vert$, the magnitude of particle
velocity quickly increases above that of the fluid point,
$\vert\vvs(t)\vert$ and eventually saturates remaining larger than
the average value of $\vert\vvs\vert$ at all times. The average
saturated particle velocity satisfies the scaling
\begin{equation}
\langle \up\rangle\sim\langle
\vs\rangle\sqrt{\ell/a_c}\,, \label{saturation2D}
\end{equation}
where $\ell$ is the intervortex spacing, and $\langle
\vs\rangle=\kappa/(2\pi\ell)$ the average superfluid velocity.

Using Schwarz's method\cite{Sch85,Sch88}, the three-dimensional
calculation of the particle motion in the vortex tangle was
performed in the periodic box. The numerical technique was
described by Samuels \emph{et al.}\cite{Sam01,K01}. The motion,
governed by Eqs.~(\ref{motionTto0}), of micron-size, neutrally
buoyant particle was studied in the statistically steady state of
the vortex tangle. In the three-dimensional case there is no need
to explicitly introduce a force-free region for distances
$\rp<a_c$ from the vortex core: in Biot-Savart calculations, the
normalization of the velocity when the solid particle approaches
too close to a vortex is achieved by the numerical cut-off of the
pressure gradient force acting on the particle, and a force-free
region is automatically provided by the discretization along the
vortex filament.

Fig.~\ref{Tr2D_Tr3D} (right) shows the typical trajectory,
projected on the $(x,\,y)$-plane, of the solid particle. This
trajectory has the same features as the two-dimensional
trajectory. Likewise, the phenomenon is observed as well of
particle velocity saturation at values of $\langle\up\rangle$ much
higher than $\langle\vs\rangle$. In the three-dimensional case the
calculated average particle velocity also agrees with
scaling~(\ref{saturation2D}).

\begin{figure}
\begin{center}
\includegraphics[%
  width=0.6\linewidth,
  keepaspectratio]{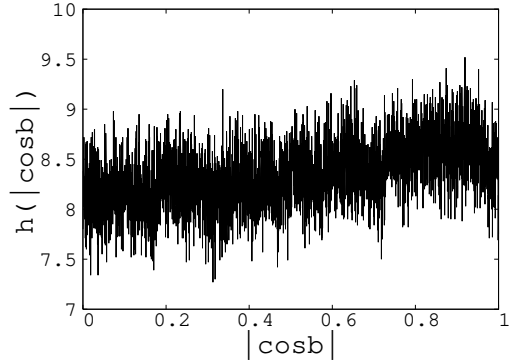}
\end{center}
\caption{Histogram\cite{K08_PoF} of the angle $b$ between $\uup$ and $\vvs$
(vertical axis is divided by 100). From Kivotides, Sergeev, and
Barenghi, \emph{Phys. Fluids}, \textbf{20},
055105, (2008). Reprinted by permission, \copyright2008 American
Institute of Physics.} \label{cos_b}
\end{figure}

These results, together with the histogram, illustrated by
Fig.~\ref{cos_b}, of the angle $b$ between $\uup$ and the
superfluid velocity $\vvs$ suggest that from the point of
view of flow visualization, in the low temperature limit (at
$T<0.5\,\textrm{K}$) the trajectories of solid particles do not
reveal flow patterns of the superfluid.

\section{Self-consistent, two-way coupling model of
particle-vortex interactions. Particle trapping on quantized
vortices} \label{self-consistent}

The problem of particle trapping on quantized vortices constitutes
a part of a wider problem of reconnections of quantized vortices
with the surface of the particle moving in the fluid velocity
field. This problem cannot be analyzed based on the one-way
coupling model introduced in Sec.~\ref{one-way}. Instead, a
more elaborate, two-way coupling model was developed by
Kivotides, Barenghi, and
Sergeev\cite{K08_PoF,K06_b,Bar07,K07_a,K08,K08_1} based on
dynamically self-consistent calculations which take into account
an influence of the flow field around the sphere on the evolution
of the superfluid vortex.

\subsection{Mathematical formulation}
\label{mathematical_formulation}

The evolution of the vortex filament represented as
a space curve $\XX(s,\,t)$, where $s$ is the arclength, is
governed by the equation
\begin{equation}
\partial\XX/\partial{t} = \VVs+ \VVb+ \VV^{\phi}+
\VV^\textrm{f}\,. \label{X}
\end{equation}
The contributions are: $\VVs$ -- Biot-Savart integral given by
formula~(\ref{Biot-Savart_1}); the potential field $\VVb$
describes the deformation of vortices due to the presence of a
stationary particle (on the particle surface
$(\VVs+\VVb)\cdot\hat{\nn}=0$, where $\hat{\nn}$ is the normal
unit vector); $\VV^{\phi}=\nabla\phi$ is the potential flow field
induced by the motion of the spherical particle:
\begin{equation}
\phi(\xx,\,t\vert\rrp)
=-\frac{1}{2}\,\frac{\ap^3}{\vert\xx-\rrp\vert^3}\uup\cdot(\xx-\rrp)\,,
\label{Vphi}
\end{equation}
where $\rrp(t)$ is the current position of the particle centre;
the contribution $\VV^\textrm{f}$ is due to the mutual friction
between the superfluid and the normal fluid:
\begin{equation}
\VV^\textrm{f}=h_{**}(\VVs+\VVb+\VV^\phi)
+h_{*}\XX'\times(\vvn-\VVs-\VVb-\VV^\phi)
+h_{**}\XX'\times(\XX'\times\vvn)\,. \label{Vf}
\end{equation}

These equations must be considered together with the equations of
motion of neutrally buoyant spherical particle; these equations
are $d\rrp/dt=\uup(t)$, and
\begin{eqnarray}
\frac{4}{3}\pi\ap^3\rhoo\,\frac{d\uup}{dt}&=&6\pi\ap\mun(\vvn-\uup)
\nonumber\\
&+&2\pi\rhos\ap^3\frac{\partial\VVs(\rrp,\,t)}{\partial
t}+\frac{1}{2}\rhos\int\limits_S
dS\,\vert\VVs+\VVb\vert^2\hat{\nn}\,, \label{particle_motion}
\end{eqnarray}
where $\rhoo$ is given by the first of relations~(\ref{rho_tau}).
At temperatures $1\,\textrm{K}<T<T_\lambda=2.17168\,\textrm{K}$,
$\mun(T)$ is the viscosity of the normal fluid. At
$T<1\,\textrm{K}$, the coefficient $\mun$ is determined by the
drag force due to ballistic scattering of quasiparticles (phonons
and rotons) off the particle surface (see the discussion later in
Sec.~\ref{temperature}).

Note that the key difference between the one-way coupling
model represented by Eqs.~(\ref{kinematic})-(\ref{motion2}) and the
self-consistent, two-way coupling model considered in this Section is
the presence of the particle-vortex interaction force represented by
the last term in the equation~(\ref{particle_motion}) of particle
motion, the latter being coupled with equations~(\ref{X})-(\ref{Vf})
governing the evolution of the vortex filament.

Numerical method of solution of the system of equations
(\ref{Biot-Savart_1}) and (\ref{X})-(\ref{particle_motion}) is
described in detail by Kivotides, Barenghi, and
Sergeev\cite{K06_b} (this method is based on generalization of the
approach by Schwarz\cite{Sch85,Sch74}, later developed further by
Tsubota and Maekawa\cite{T93}, to the problem of reconnection of
the quantized vortex with the stationary surface or the surface
moving with prescribed velocity).

\begin{figure}
\begin{center}
\includegraphics[%
  width=0.65\linewidth,
  keepaspectratio]{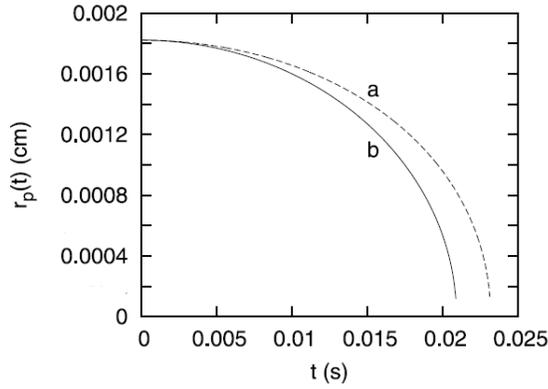}
\end{center}
\caption{Distance between the particle and the vortex core vs time:
(a) one-way coupling model; (b) self-consistent two-way coupling
model\cite{Bar07}. From Barenghi, Kivotides, and Sergeev, \emph{J. Low
Temp. Phys.}, \textbf{148}, 293, (2007).
Reprinted by permission, \copyright2007 Springer.}
\label{distance}
\end{figure}

Using this model we may analyze first how incorrect is the
one-way coupling model. Fig.~\ref{distance} shows the
distance between the particle and the initially straight vortex vs
time found from (a) analytical calculation\cite{P05} based on the
one-way coupling model, and (b) numerical, dynamically
self-consistent, two-way coupling model\cite{Bar07} given by
Eqs.~(\ref{Biot-Savart_1}), (\ref{X})-(\ref{particle_motion}). As
can be seen, despite the simplicity of the one-way coupling model,
agreement is good until the moment when the particle and the
vortex are so close that the vortex reconnects to the particle
surface.

\subsection{Mechanism of particle-vortex interaction}
\label{mechanism_interaction}

We will focus here only on the most important aspects of
particle-vortex interaction, referring the interested reader to
original publications of Kivotides, Barenghi, and
Sergeev\cite{K08_PoF,K06_b,Bar07,K07_a,K08,K08_1}.

All the calculations illustrated below were performed for the
neutrally buoyant particle of radius $\ap=1\,\mu\textrm{m}$
located initially at the distance $2\ap$ from the initially
straight vortex filament. To simplify the analysis, in all
examples considered below it was assumed that
$\vvn\equiv\mathbf{0}$.

\begin{figure}[t]
\begin{tabular}[b]{ccc}
\includegraphics[height=0.30\linewidth]{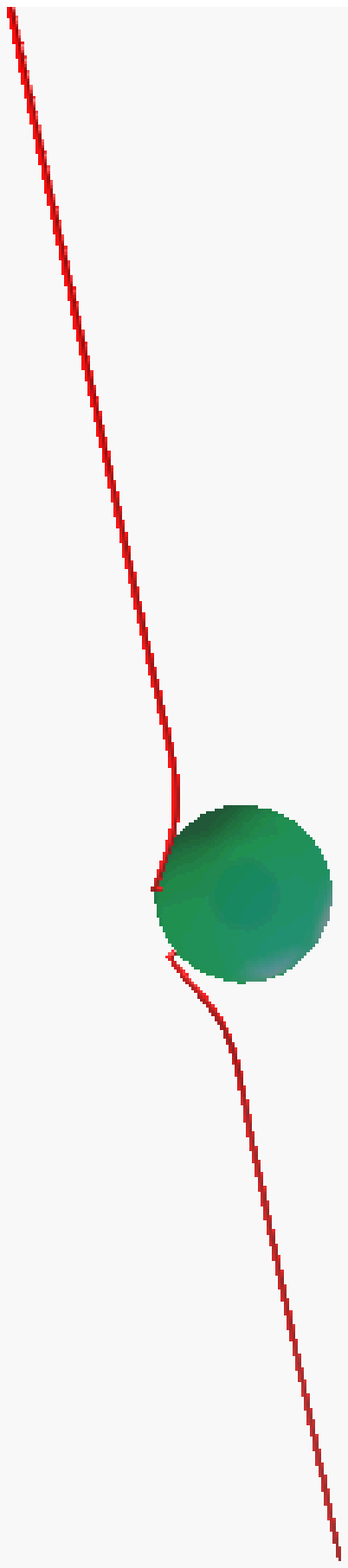}&
\includegraphics[height=0.30\linewidth]{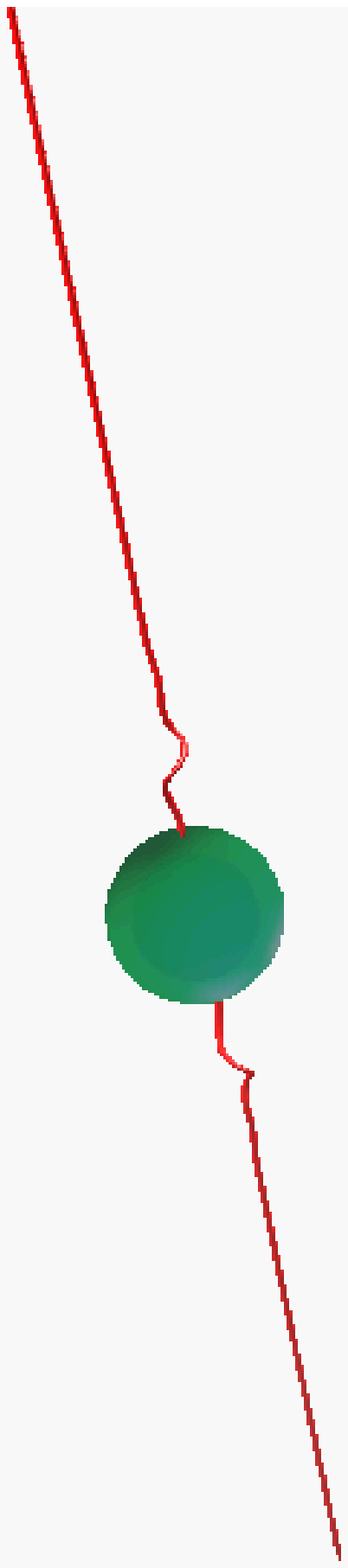}&
\includegraphics[height=0.30\linewidth]{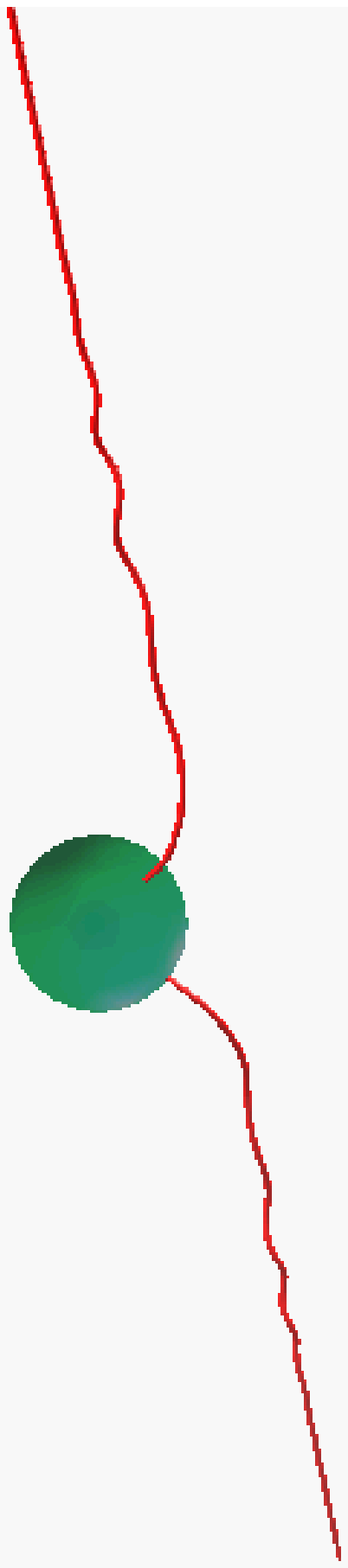}\cr
\includegraphics[height=0.30\linewidth]{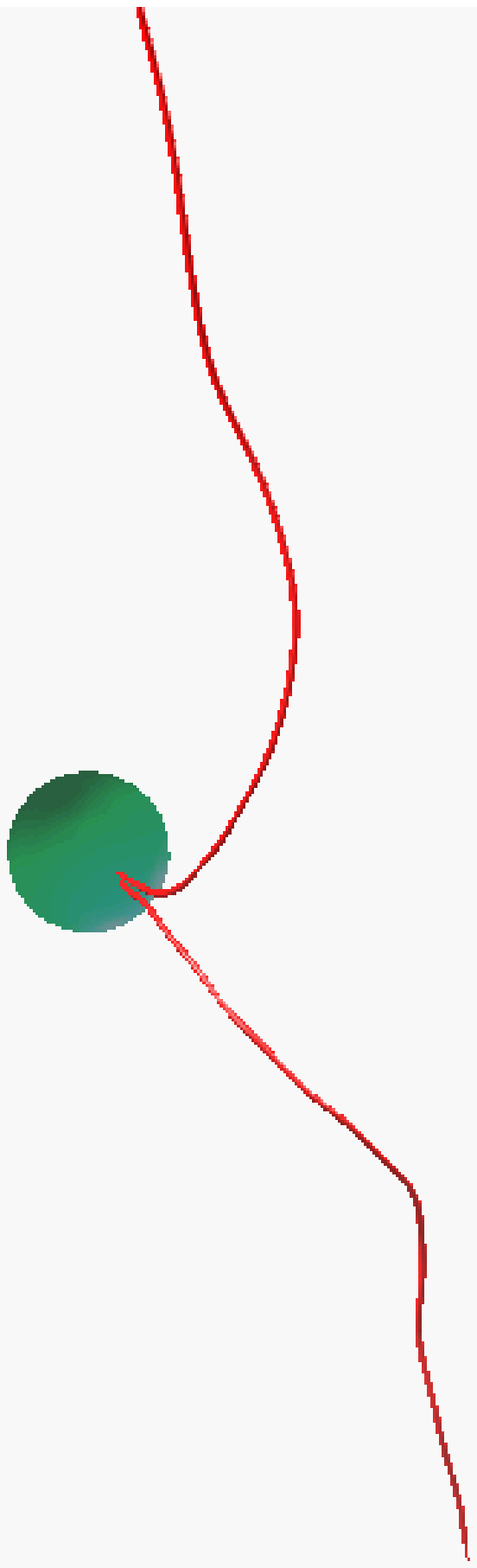}&
\includegraphics[height=0.30\linewidth]{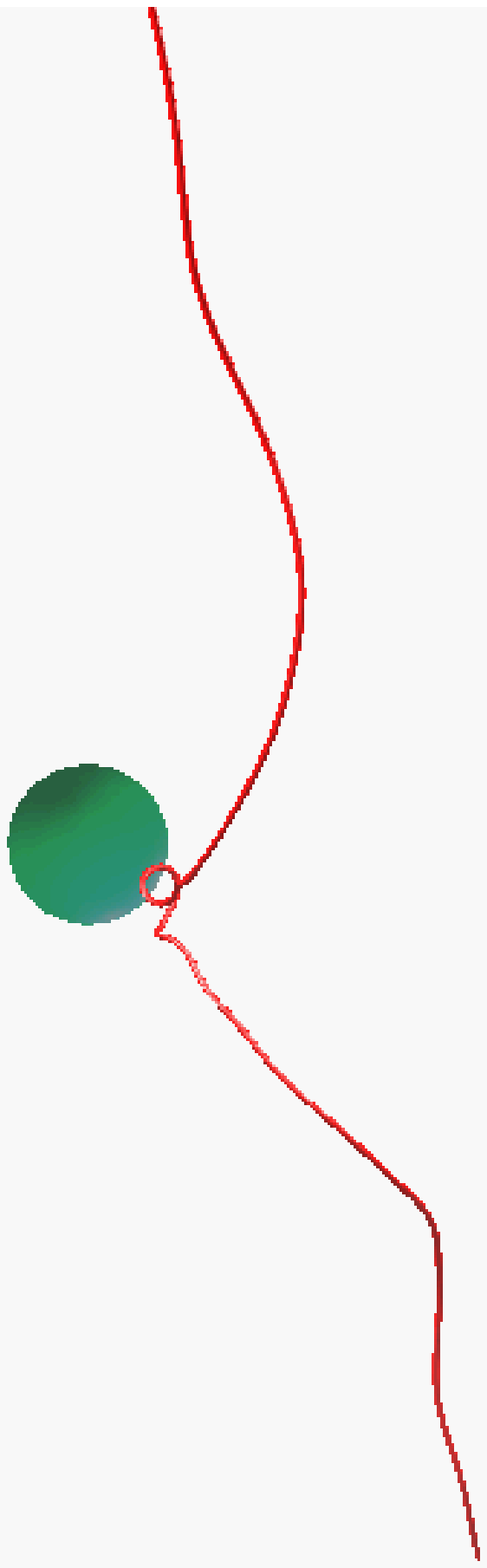}&
\includegraphics[height=0.30\linewidth]{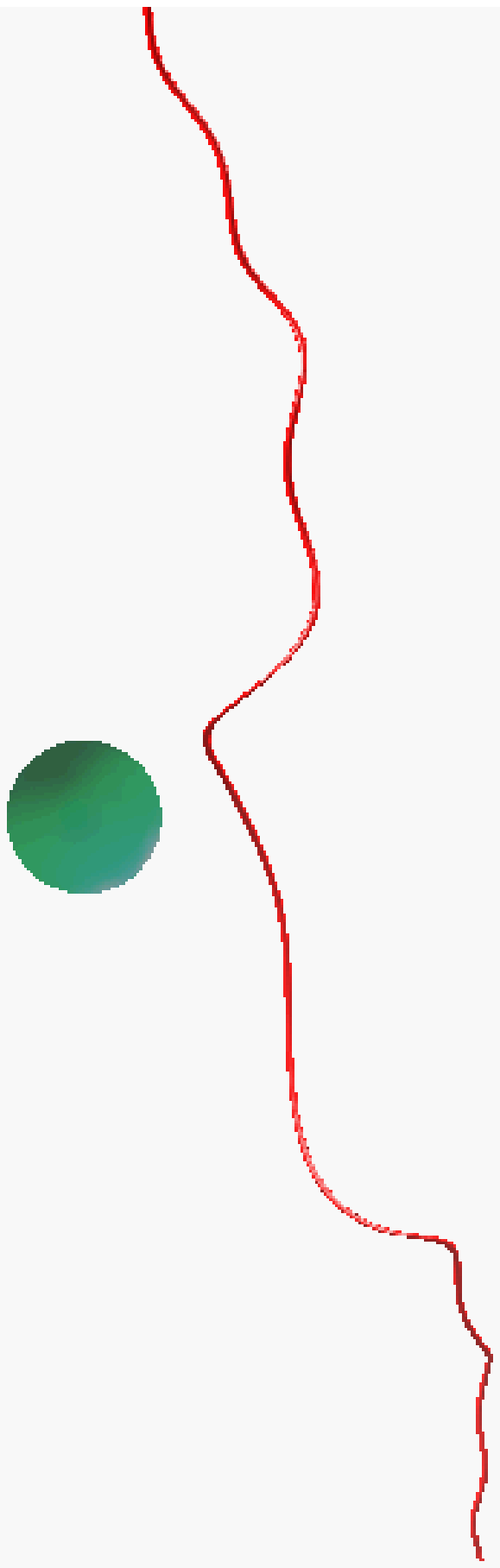}\\
\end{tabular}
\caption{(Color online) Particle-vortex collision\cite{K08} at
$T=1.3\,\textrm{K}$.
Initial velocity $\up=25\,\textrm{cm}/\textrm{s}$. Particle moves
from the right. From Kivotides, Barenghi, and Sergeev, \emph{Phys. Rev.
B}, \textbf{77}, 014527, (2008).
Reprinted by permission, \copyright2008 American Physical
Society.} \label{collision_T1.3}
\end{figure}

We illustrate first an influence of the initial velocity of solid
particle on the particle-vortex collision\cite{K08}.
Fig.~\ref{collision_T1.3} shows a sequence of particle-vortex
configurations at $T=1.3\,\textrm{K}$ for initial velocity
$\up=25\,\textrm{cm}/\textrm{s}$. The particle arrives from the
right; \emph{top left}: the vortex is deformed as it tries to
avoid the incoming particle; \emph{top middle and top right}: the
reconnection of vortex to the particle surface excites Kelvin
waves; \emph{bottom left}: the particle drags the vortex, forcing
its two strands to come together, thus facilitating a second
reconnection; \emph{bottom middle}: following the reconnection,
the vortex recoils and the particle breaks free; \emph{bottom
right}: the following relaxation creates more Kelvin waves.

\begin{figure}[t]
\begin{tabular}[b]{ccc}
\includegraphics[height=0.30\linewidth]{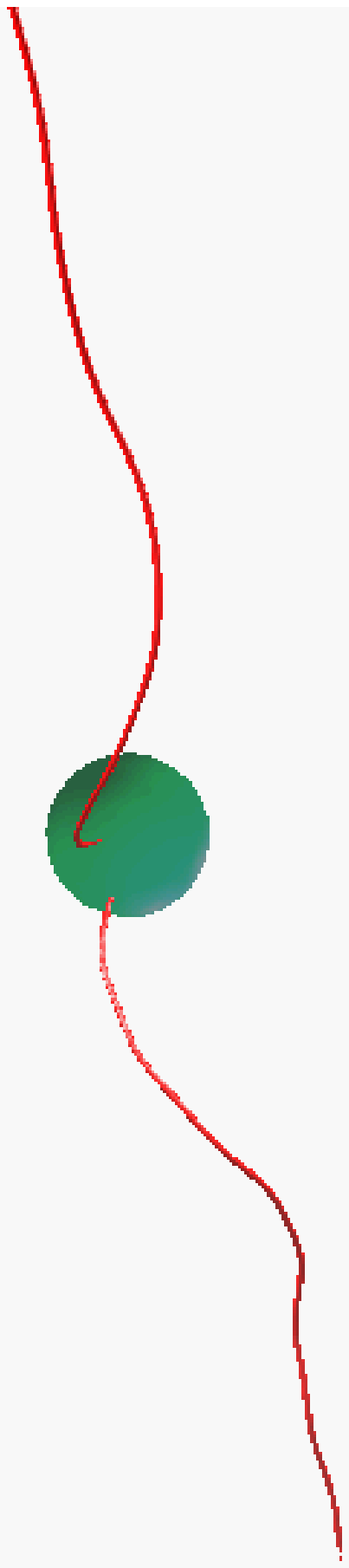}&
\includegraphics[height=0.30\linewidth]{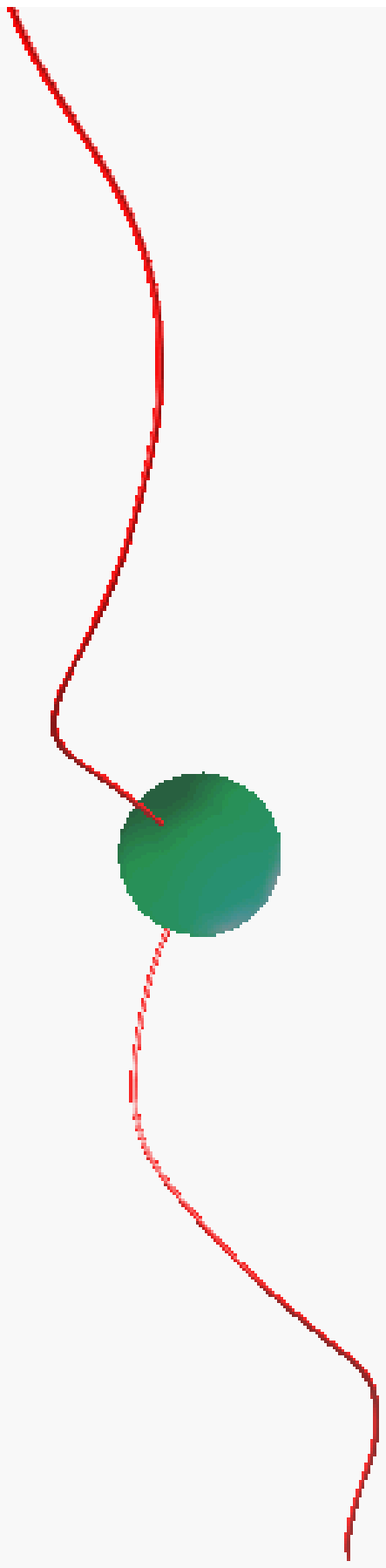}&
\includegraphics[height=0.30\linewidth]{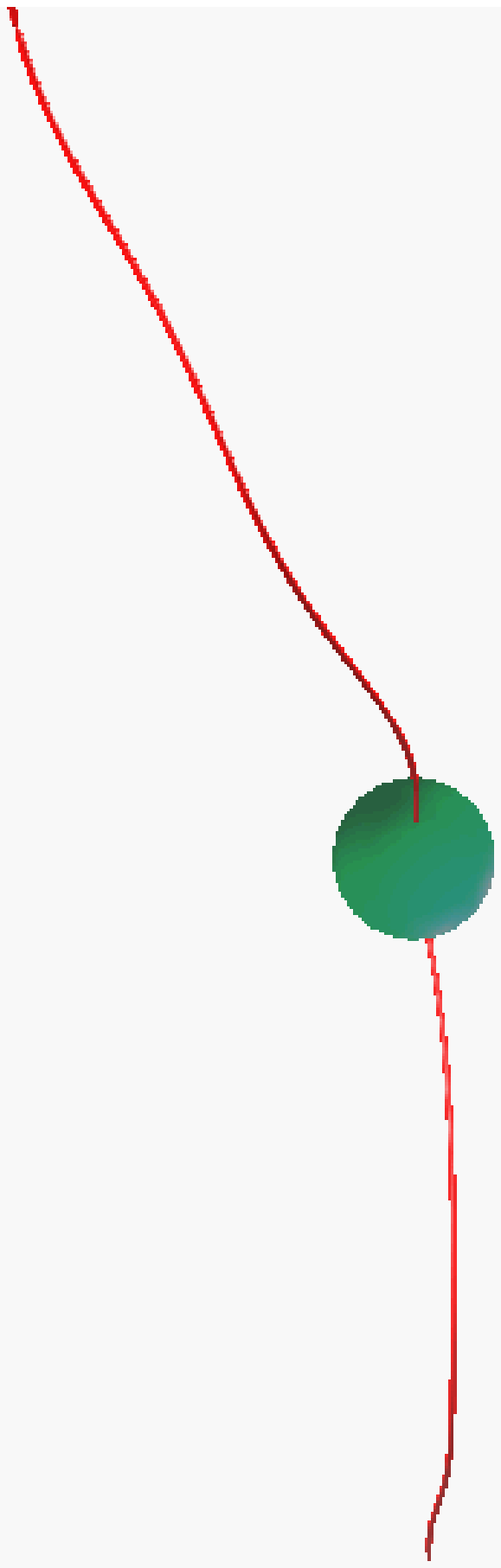}\\
\end{tabular}
\caption{(Color online) Particle trapping on the vortex core\cite{K08}.
$T=1.3\,\textrm{K}$, initial velocity
$\up=20\,\textrm{cm}/\textrm{s}$. From Kivotides, Barenghi, and Sergeev,
\emph{Phys. Rev.
B}, \textbf{77}, 014527, (2008).
Reprinted by permission, \copyright2008 American
Physical Society.} \label{trapping_T1.3}
\end{figure}

The second computation is carried out at the same temperature for
the smaller initial velocity $\up=20\,\textrm{cm}/\textrm{s}$. In this
case, illustrated by Fig.~\ref{trapping_T1.3}, the particle is
trapped by the vortex: after reconnecting with the vortex the
particle slows down and stops since it lacks the kinetic energy to
stretch the vortex and to induce the second reconnection. Results
illustrated by Figs.~\ref{collision_T1.3} and \ref{trapping_T1.3}
indicate that, for each temperature, there exists a
critical velocity, $v_{cr}$ of the particle-vortex approach;
provided the relative velocity of the particle and the vortex is
smaller than $v_{cr}$, the particle will be trapped on the
quantized vortex core.

Note that in these two, somewhat artificial examples a
possibility of nucleation of quantized vortices (and hence extra
dissipation) by a moving particle has been ignored. A rather high
particle velocities (20 and $25\,\textrm{cm/s}$) were used for the
purpose of illustration only; similar results were obtained as well
for considerably smaller velocities. It is unlikely that in the real
turbulent $^4$He the particle velocity relative to thge vortex core 
can be as high as in these illustrations. For example, in the
counterflow turbulence $\up$ can hardly be larger than $v_{ns}$, the
latter usually being considerably smaller than the critical nucleation
velocity.

It was also found\cite{K07_a} that having been trapped by the
quantized vortex the particle may then drift along the vortex
filament. For a neutrally buoyant, micron-size particle, a typical
drift velocity, $v_{drift}$ was found to be about
$0.5\,\textrm{cm}/\textrm{s}$. Such a drift can be explained by
the interaction of the particle with Kelvin waves (which are not
necessarily symmetric with respect to the particle) induced by the
particle-vortex collision. The drift velocity provides a rather
simple way of estimating the amplitude of collision-induced Kelvin
waves, e.g. by modeling a Kelvin wave of amplitude $A$ as a vortex
ring of radius $A$ and then balancing the momentum of the ring
with that of the drifting particle. For
$v_{drift}\approx0.5\,\textrm{cm}/\textrm{s}$ such a procedure
yields $A\approx0.25\,\mu\textrm{m}$.

\subsection{Influence of temperature on particle-vortex
collision and particle trapping.} \label{temperature}

\begin{figure}
\begin{center}
\includegraphics[%
  width=0.95\linewidth,
  keepaspectratio]{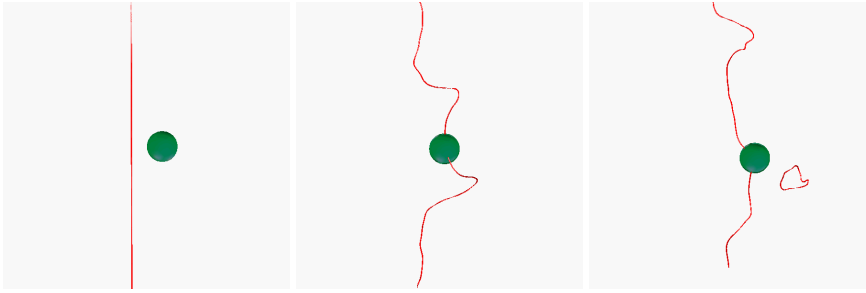}
\end{center}
\caption{(Color online) Particle trapping on the vortex core\cite{K08_PoF} at
$T=1.3\,\textrm{K}$ and $\up=0$. From Kivotides, Sergeev, and Barenghi,
\emph{Phys. Fluids}, \textbf{20},
055105, (2008). Reprinted by permission,
\copyright2008 American Institute of Physics.}
\label{trapping_emission}
\end{figure}

Another computation\cite{K08_PoF} was carried out at temperature
$T=1.3\,\textrm{K}$ for the particle initially at rest. The
particle starts moving under the influence of the radial pressure
gradient generated by the vortex. Fig.~\ref{trapping_emission}
shows the sequence of particle-vortex configurations at times
$t=0$ (left), $0.125\times10^{-2}\,\textrm{s}$ (center), and
$0.263\times10^{-2}\,\textrm{s}$ (right). The last frame shows
that the vortex traps the particle and emits a small vortex ring
(note that this feature is not common for all trapping events)
thus reducing the total energy of the particle-vortex
configuration. The results of calculation\cite{K08_PoF} for the
same initial configuration and particle velocity, but $\mun$
assumed to be only 0.2 of its value at $T=1.3\,\textrm{K}$
seem to suggest
that at any, however small, non-zero initial velocity, the particle,
although undergoing the process of reconnection with the vortex
filament, eventually breaks free, so that trapping does not occur.
Similar (in fact, almost identical) scenario is typical of $T\to0$
limit when the (viscous) damping force acting on the particle can
be neglected. (However, this issue is less trivial than it seems
-- see the discussion below.)

These examples show that the the presence of the damping is
crucial for trapping of solid particles by quantized vortices, so
that it can be, rather naively, expected that trapping cannot
occur at temperatures below $1\,\textrm{K}$ when the normal fluid
is absent. However, at $T<1\,\textrm{K}$ there still exists a
damping force, $ \FF^\textrm{d}=-6\pi\ap\mun\uup$ caused by
ballistic scattering of quasiparticles (phonons and rotons) off
the particle surface. Note the Stokesian form of this force (cf.
Eq.~(\ref{Stokes}), although $\mun(T)$ should now be understood
not as a viscosity but as the damping coefficient).

\begin{figure}
\begin{center}
\includegraphics[%
  width=0.65\linewidth,
  keepaspectratio]{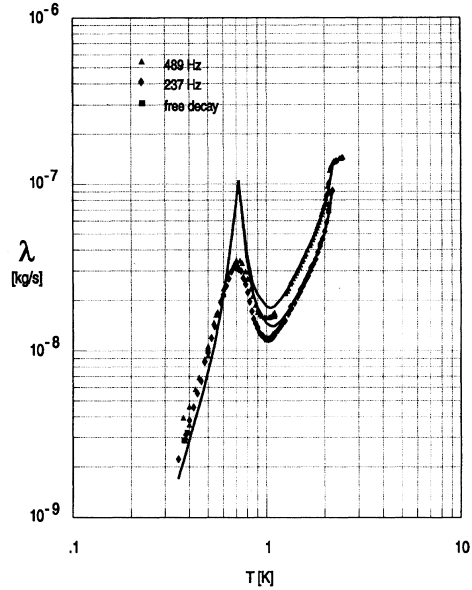}
\end{center}
\caption{Damping coefficient $\lambda$ as a function
of temperature\cite{J95} (experimental results of J\"ager, Schuderer, and Schoepe,
\emph{Phys. Rev. Lett.}, \textbf{74}, 566, (1995)).
Reprinted by permission, \copyright1995 American Physical Society.}
\label{Schoepe}
\end{figure}

Calculations\cite{K08_PoF} for various values of the damping
coefficient showed that trapping does not occur in the case where
the damping coefficient, $\mun$ is smaller than
$0.2\times\mun(1.3\,\textrm{K})$. The value of the damping
coefficient was measured experimentally by J\"ager,
Schuderer, and Schoepe\cite{J95} for the spherical particle
of radius $100\,\mu\textrm{m}$. Fig.~\ref{Schoepe} shows the
experimental results\cite{J95}
for the coefficient $\lambda=6\pi\ap\mun$ in the range of
temperatures including both the region $T>1\,\textrm{K}$
corresponding to the classical viscous dissipation in the normal
fluid, and $T<1\,\textrm{K}$ corresponding to the regime of
ballistic scattering of quasiparticles. It can be seen that in the
temperature interval $0.6\,\textrm{K}<T<1\,\textrm{K}$ the value
of $\mun$ may even exceed the viscosity at temperatures above
$1\,\textrm{K}$. Only at temperatures below $0.5\,\textrm{K}$ the
damping coefficient becomes smaller than
$0.2\times\mun(1.3\,\textrm{K})$. Therefore, it can be 
expected that the trapping of neutrally buoyant,
$1\,\mu\textrm{m}$ particles does not occur at temperatures below
$0.5\,\textrm{K}$.

At these, very low temperatures the motion of solid particle can
be modeled by a simpler, one-way coupling model which is based on
the assumption that particles are not trapped on quantized vortex
lines. However, at these temperatures the presence and motion of
the particles still affects the vortex filaments, so that a
certain modification should be necessary of the one-way
coupling model. Such a modification was already discussed above in
Sec.~\ref{Tto0}.

It was already mentioned rather briefly that the results
discussed above can be invalidated in the case where the extra
dissipation is produced by the nucleation of quantized vortices in the
vicinity of the particle whose velocity relative to the superfluid
component is sufficiently large. The details of this mechanism are not
properly understood yet and require further, rather complicated 
numerical study. Perhaps the only example of such a study so far is
the work of H\"anninen, Tsubota, and Vinen\cite{Hanninen} who hinted
at the possibility of formation of a wake, growing with time, of
quantized vorticity behind an oscillating sphere. However, we
anticipate that relatively small particle velocities typical of PIV or
particle tracking experiments allow to ignore the extra dissipation 
due to the nucleation of quantized vortices.

\subsection{Self-consistent model of particle collisions with
vortex rings} \label{Kivotides_Wilkin}

\begin{figure}
\begin{center}
\includegraphics[%
  width=0.99\linewidth,
  keepaspectratio]{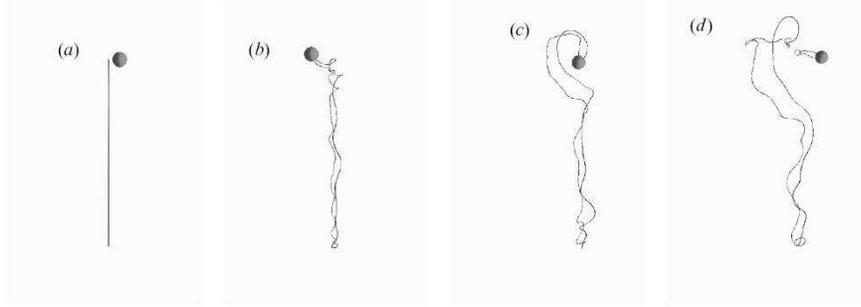}
\end{center}
\caption{Example\cite{K_Wil08} of the particle-vortex interaction at $T=0$.
Initial radius and velocity of the ring are
$R=1.25\times10^{-3}\,\textrm{cm}$ and
$V_R=0.771\,\textrm{cm}/\textrm{s}$. The particle is initially at
rest. Four frames, from left to right, correspond to times $t=0$,
$3.407\times10^{-4}$, $5.679\times10^{-4}$, and
$7.698\times10^{-4}\,\textrm{s}$.
From Kivotides and Wilkin, \emph{J. Fluid Mech.}, \textbf{605},
pp.~367-387, (2008). \copyright2008 Cambridge
University Press.} \label{particle_vortex}
\end{figure}

The self-consistent, two-way coupling model described in this
Section, was recently applied by Kivotides and
Wilkin\cite{K_Wil08} for numerical study of interactions between
neutrally buoyant solid particles and quantized vortex rings in
the range of temperatures between $T=0$ and $T=T_\lambda$. It was
found that trapping of particles by sufficiently small vortex rings never occurs, and
that, at $T=0$, the dominant dynamical process in the
particle-ring interaction is the excitation and propagation of
Kelvin waves along the vortex ring. The collision between the
particle initially at rest and the vortex ring of radius
$1.25\times10^{-3}\,\textrm{cm}$ at temperature $T=0$ is
illustrated by the sequence shown in Fig.~\ref{particle_vortex}.
It was found that typical of the particle-vortex collisions at
$T=0$ is spiraling of the particle out of the point of initial
contact with the vortex ring. At finite temperatures the
particle-vortex collision induces particle oscillations in the
direction normal to the particle trajectory. As should be
expected, at finite temperatures the mutual friction damps the
amplitude and reduces the frequency of Kelvin waves propagating
along the ring.

\section{Visualization experiments and their theoretical interpretation}
\label{PIVexperiments}

\subsection{Particle motion in turbulent thermal counterflow}
\label{VanSciver_counterflow}

\subsubsection{Experiment} \label{experiment_counterflow}

One of the first and, in our view, one of the most important
experiments illustrating strong interactions between solid
particles and quantized vortices was performed by Zhang and Van
Sciver\cite{Z05_1} who studied the sedimentation of heavy
($\rhop=1.1\,\textrm{g}/\textrm{cm}^3$) particles in thermal
counterflow produced by the heat source situated at the bottom of
vertical apparatus, so that the normal fluid flows upwards. The experiments were
performed in the temperature range from 1.62 to $2.0\,\textrm{K}$; the applied heat
flux ranged from 110 to $1370\,\textrm{mW}/\textrm{cm}^2$.

\begin{figure}
\begin{center}
\includegraphics[%
  width=0.65\linewidth,
  keepaspectratio]{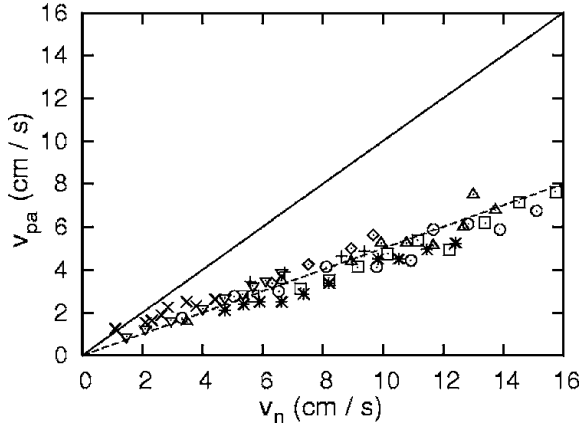}
\end{center}
\caption{PIV measurement of Zhang and Van Sciver\cite{Z05_1} of
$v_{pa}=\up+v_{slip}$ for particles sedimenting in
turbulent thermal counterflow. Solid line: $v_{pa}$ calculated according to
Eqs.~(\ref{vn_vpa}) and (\ref{v_slip}); dashed line is calculated\cite{S06_b} from
Eqs.~(\ref{vpa}) and (\ref{v_add}) with $\beta=3$. Reprinted, by permission,
from Sergeev, Barenghi, and Kivotides, \emph{Phys. Rev.
B}, \textbf{74}, 184506, (2006).
\copyright2006 American Physical Society.} \label{PIV_sedimentation}
\end{figure}

It
seemed natural to expect that the dominating force acting on the
particle will be the viscous drag force and, therefore, the
particle velocity will be
\begin{equation}
\up=\vn-v_{slip}\,, \label{vn_vpa}
\end{equation}
where $v_{slip}$ is the terminal velocity of particle
sedimentation given by relation
\begin{equation}
v_{slip}=\frac{2\ap^2}{9\mun}\,(\rhop-\rho)\,. \label{v_slip}
\end{equation}
Were this correct,
the experimental data for $v_{pa}=\up+v_{slip}$
plotted against $\vn$ would have collapsed on the straight solid line
shown in Fig.~\ref{PIV_sedimentation}.
However, as can be seen
from this Fig.~\ref{PIV_sedimentation}, the results of PIV measurement showed much lower
particle velocity. Zhang and Van Sciver\cite{Z05_1} found that, instead of
(\ref{v_slip}), the particle velocity can be represented as
$\up=\vn-v_{slip}-v_{add}$, where the additional
velocity, $v_{add}$ can be explained only by strong
interactions between sedimenting particles and the vortex tangle.
Zhang and Van Sciver also found that $v_{pa}/\vn\approx0.5$
independently of temperature, and $v_{add}\sim q=\rho
ST\vn$.

\subsubsection{Phenomenological theory of particle motion} \label{phenomenology}

Below we will discuss the phenomenological theory, developed by
Sergeev, Barenghi and Kivotides\cite{S06_b}, 
of the motion of micron-size particles in thermal
counterflow and hence explain the surprising result of Zhang and
Van Sciver's experiment.  The relatively simple
analytical model  of Sergeev, Barenghi
and Kivotides arises from the physlcal insight acquired using the numerical
(and computationally expensive) two-way coupling model.

Imagine that two strands of the quantized vortex are attached to
the surface of spherical particle as shown in
Fig.~\ref{asymmetric} (left) (at the
point of reconnection the vortex strand is necessarily orthogonal
to the particle surface). The force exerted by a vortex strand
attached to the surface is $\FF=\int_S p\,\hat{\nn}\,dS$ which, as
shown by Schwarz\cite{Sch74}, can be written as
\begin{equation}
\FF=\frac{\rhos}{2}\int\limits_S\vert\VVs+\VVb\vert^2\hat{\nn}\,dS\,.
\label{Schwarz_force}
\end{equation}
The contribution of $\VVb$ to
this force can be neglected in the case where the radius of
curvature of the vortex strand is much larger than the particle
radius.

The vortex tangle in the counterflow can be so dense that several
vortex strand can be simultaneously attached to the particle. Since the quantum of
circulation is small, the leading contribution to the
integral~(\ref{Schwarz_force}) is provided by a small area around
the point where the vortex attaches to the surface. This enables
us to find the following analytic approximation for the force
exerted on the particle:
\begin{equation}
\FF\approx\frac{\rhos\kappa^2}{4\pi}\ln\frac{\ap}{\xi}\sum_{i=1}^N\hat{\nn}_i\,.
\label{F_analytic}
\end{equation}
It can be noticed that, in agreement with the experimental results\cite{Z05_1},
$\FF$ is a body force.

\begin{figure}
\begin{center}
\includegraphics[%
  width=0.70\linewidth,
  keepaspectratio]{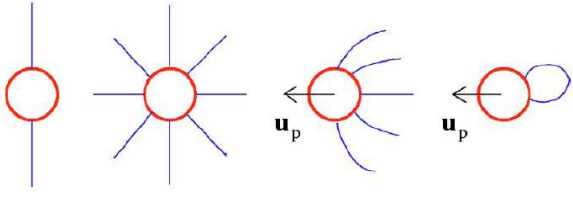}
\end{center}
\caption{(Color online) Possible particle-vortex configurations\cite{S06_b}.
From Sergeev, Barenghi, and Kivotides, \emph{Phys. Rev.
B}, \textbf{75}, 019904(E), (2007).
Reprinted by permission, \copyright2007 American Physical Society.}
\label{configurations}
\end{figure}

Several possible particle-vortex configurations shown in
Fig.~\ref{configurations} can be imagined. If configuration is
symmetric, as in the first two figures, the net force is zero. If
one or more vortex loops are asymmetrically attached to the
sphere, as in the third and the fourth figures, the contributions
from individual vortices will not cancel out resulting in a net
body force.

The following scenario seems realistic: the sphere, as it moves
between vortex lines, carries along one or more vortex lines or
even separate loops as the result of previous close encounter with
vortices. This scenario corresponds to asymmetric particle-vortex
configurations, shown in
Fig.~\ref{configurations}, leading
to emergence of the force exerted on the particle in the direction
opposite to its motion. In the case of moderately dense
tangle ($\ap\lesssim\ell$) typical of the experiment of Zhang and Van
Sciver\cite{Z05_1},
the average force exerted on the particle by the vortex tangle can
be calculated\cite{S06_b} as
\begin{equation}
F\approx\frac{\rhos\kappa^2}{4\pi}\,
\biggl(\frac{2\beta\ap}{\ell}\biggr)\,\ln\frac{\ap}{\xi}\,.
\label{F_dilute}
\end{equation}
Here the only unknown quantity is the parameter
$\beta$ which should be determined by geometrical properties of
the vortex tangle in the vicinity of the particle. It can be
expected that $\beta=O(1)$.

The additional velocity can now be calculated as
$v_{add}=F/(6\pi\ap\mun)$, so that for
$v_{pa}=\up+v_{slip}$ we find
\begin{equation}
v_{pa}=\vn-v_{add}
\approx\biggl(1-\frac{\beta\rho\kappa^2\gamma\ln(\ap/\xi)}
{12\pi^2\mun}\biggr)\,\vn\,, \label{vpa}
\end{equation}
where $\gamma(T)=L^{1/2}/v_{ns}=\rhos L^{1/2}/(\rho\vn)$ is
the known mutual friction coefficient, with $v_{ns}=\vn-\vs$. In agreement with the
experimental results\cite{Z05_1} we find that $v_{pa}$
is proportional to the normal fluid velocity. Moreover, the
temperature dependence of the slope $v_{pa}/\vn$ turns out
to be the same as in the cited experiment.

Using formula~(\ref{vpa}), the ratio of the additional velocity to
the heat flux can be calculated as a function of temperature:
\begin{equation}
\frac{v_{add}}{q}
=\frac{\beta\kappa^2\ln(\ap/\xi)}{12\pi^2}\biggr(\frac{\gamma}{\mun
ST}\biggr)\,, \label{v_add}
\end{equation}
so that $v_{pa}$ can now be calculated
explicitly. The dashed line in Fig.~\ref{PIV_sedimentation} reproduces $v_{pa}$
calculated by Sergeev, Barenghi, and Kivotides\cite{S06_b} using
formulae~(\ref{vpa})-(\ref{v_add}) with
$\beta=3$. As can be seen, the developed phenomenological theory
agrees well, not only qualitatively but also quantitatively, with
PIV measurements of Zhang and Van Sciver and, therefore, 
seems to explain
the mechanism of particle motion in the thermal counterflow.

Calculations similar to those leading to
formulae~(\ref{F_dilute})-(\ref{v_add}) can also be performed in
the limit of the very dense vortex tangle such that $\ap\gg\ell$. In this
case reconnections between the particle and vortices happen all
the time, so that the particle is always attached to several vortex
filaments. Calculation of the average force exerted on the
particle by the vortex tangle yields\cite{S06_b}:
\begin{equation}
F\approx\frac{\rhos\kappa^2}{4\pi}\,2\beta_d
\biggl(\frac{\ap}{\ell}\biggr)^2\ln\frac{\ap}{\xi}\,,
\label{F_dense}
\end{equation}
where $\beta_d=O(1)$ is again a geometrical factor, and
$(\ap/\ell)^2$ represents the cross-section of the particle
interaction with the network of vortices. Proceeding as in the previous case,
a different (cf.
Eq.~(\ref{vpa})) dependence of $v_{pa}$ on $\vn$, $T$
and $\ap$ is predicted:
\begin{equation}
v_{pa}\approx\vn\biggl[1-\frac{\beta_d\ap(\kappa\gamma\rho)^2\ln(\ap/\xi)}{12\pi^2\mun\rhos
f}\,\vn\biggr]\,, \label{vpa_dense}
\end{equation}
where $f=1+\beta_d\gamma/(3\pi\mun)$.
Finding whether or not this prediction agrees with observations
would require a new experiment similar to that of Zhang and Van Sciver\cite{Z05_1} but
for considerably higher values of the vortex line density (i.e.
such that $L\gg\ap^{-2}$).

\subsubsection{Self-consistent model of particle motion in the thermal counterflow}
\label{self_consistent_counterflow}

In order to justify the phenomenological theory of
Sec.~\ref{phenomenology} and investigate the particle motion in
more detail, Kivotides\cite{K08_a,K08_b} applied the two-way
coupling, self-consistent model described in
Sec.~\ref{mathematical_formulation} for a numerical study of
particle interactions with the vortex tangle in thermal
counterflow. Although in the cited works the motion of not a heavy
but neutrally buoyant particle was studied, the results reveal
some important aspects of the experiments performed by Zhang and
Van Sciver~\cite{Z05_1} and Paoletti \emph{et al.}\cite{P08_a}.

We start this Section with reviewing the first of these papers. In
calculations\cite{K08_a} the normal fluid velocity in the counterflow was assumed a
constant, prescribed value (this was the only realistic option considering the
complexity of the model represented by Eqs.~(\ref{X})-(\ref{particle_motion})).
Numerical analysis of the particle motion was performed for the statistically steady
vortex tangle modeled in the same way as in the earlier work~\cite{K06} described in
Sec.~\ref{counterflow_one_way}.

Calculations were performed for three temperatures,
$T=1.3$, 1.95, and $2.171\,\textrm{K}$.
It was found that the following four factors strongly affect
the particle motion: 1)~stratification of the vortex tangle,
2)~vortex line density, 3)~the average drift of the tangle,
and 4)~the intensity of Kelvin wave cascades,
induced by particle-vortex collisions, along vortex filaments.
For example, for $T=1.95\,\textrm{K}$ Kivotides found that
the tangle is strongly stratified, and an average drift of the tangle
is small. Since the vortices expand mostly in the direction normal
to that of the counterflow, it was also found that in the
considered case the stratification does not affect the average
properties of the particle motion. The statistically steady
particle motion is governed by the balance of the Stokes force acting
in the direction of the normal flow, and the average particle-vortex
interaction force acting in the opposite direction; the latter force
was found to be proportional to the vortex line density and the mean
relative particle-tangle velocity.

The calculations\cite{K08_a} confirmed the mechanism suggested by the
phenomenological theory\cite{S06_b} discussed in the previous Section: a formation
of vortex loops attached to the rear part of the particle surface causes an
additional force opposite to the direction of particle motion and hence reduces the
slip velocity. For $T=1.95\,\textrm{K}$ and the vortex line density
$L=3.284\times10^7\,\textrm{cm}^{-2}$ Kivotides' calculation yielded
$\langle\up\rangle\approx0.6\vn$, in a very good agreement with experimental results
of Zhang and Van Sciver\cite{Z05_1}.

At the same temperature but higher vortex line density,
$L=8.284\times10^7\,\textrm{cm}^{-2}$ Kivotides found that the ``head-on''
particle-vortex collisions are more important and counterbalance the force caused by
the formation
of vortex loops in the rear part of the particle, so that
$\langle\up\rangle\approx\vn$. It was also found that the drift of the tangle in the
direction opposite to the particle velocity increases the frequency of the
``head-on'' particle-vortex collisions and hence the average particle velocity. This
effect becomes more pronounced at higher temperature; thus, for
$T=2.171\,\textrm{K}$ and $L=5.846\times10^7\,\textrm{cm}^{-2}$ it was found that
$\langle\up\rangle\approx1.2\vn$, again in a very good agreement with experimental
results\cite{Z05_1}.

At lower temperature, $T=1.3\,\textrm{K}$, Kelvin waves generated by particle-vortex
collisions decay slower than at higher temperatures. This, in turn, leads to large
amplitudes of particle velocity fluctuations which become comparable with
$\langle\up\rangle$. The intensity of these fluctuations was found to be inversely
proportional to the temperature and the average velocity of the tangle drift
relatively to the particle.

As argued 
in Sec.~\ref{mechanism_interaction} (for details see original publications
of Kivotides, Barenghi, and Sergeev\cite{K06_b,K08}), at
$0.5\,\textrm{K}<T<T_\lambda$ there should exist a critical velocity of the
particle-vortex approach below which a neutrally buoyant particle will necessarily
be trapped by the initially straight quantized vortex. Kivotides noted~\cite{K08_b}
that at high counterflow velocities the tangle is very dense and, therefore,
particle-vortex interactions become so strong that it may no longer be possible to
measure the normal velocity. In the cited work he estimated
the parameters of counterflow in which the normal velocity would be above the
trapping limit but yet sufficiently low, such that the tangle is sufficiently dilute
to enable the measurement of the normal velocity by the PIV or the particle tracking
technique.

Using the same model as in his previous work\cite{K08_a}, Kivotides\cite{K08_b}
analyzed numerically, at temperature $T=1.3\textrm{K}$, the particle motion in the
thermal counterflow with the normal velocity $\vn=10\,\textrm{cm}/\textrm{s}$ and
the vortex line density $L=1.168\times10^6\,\textrm{cm}^{-2}$. Performing
calculations for different initial positions of the particle, Kivotides found that
in more than 50\% of realizations the particle moved through the computational
domain without experiencing any collision with the vortex tangle. On average, the
deviation, caused by particle-vortex collisions, of the particle velocity from
that of the normal fluid was found to be less than 4\%. Moreover, in the case where
the particle does not follow the normal fluid it tracks the motion of the vortex
tangle. On the time series of particle velocity the events of the particle-vortex
\begin{figure}[t]
\begin{tabular}[b]{cc}
\includegraphics[height=0.28\linewidth]{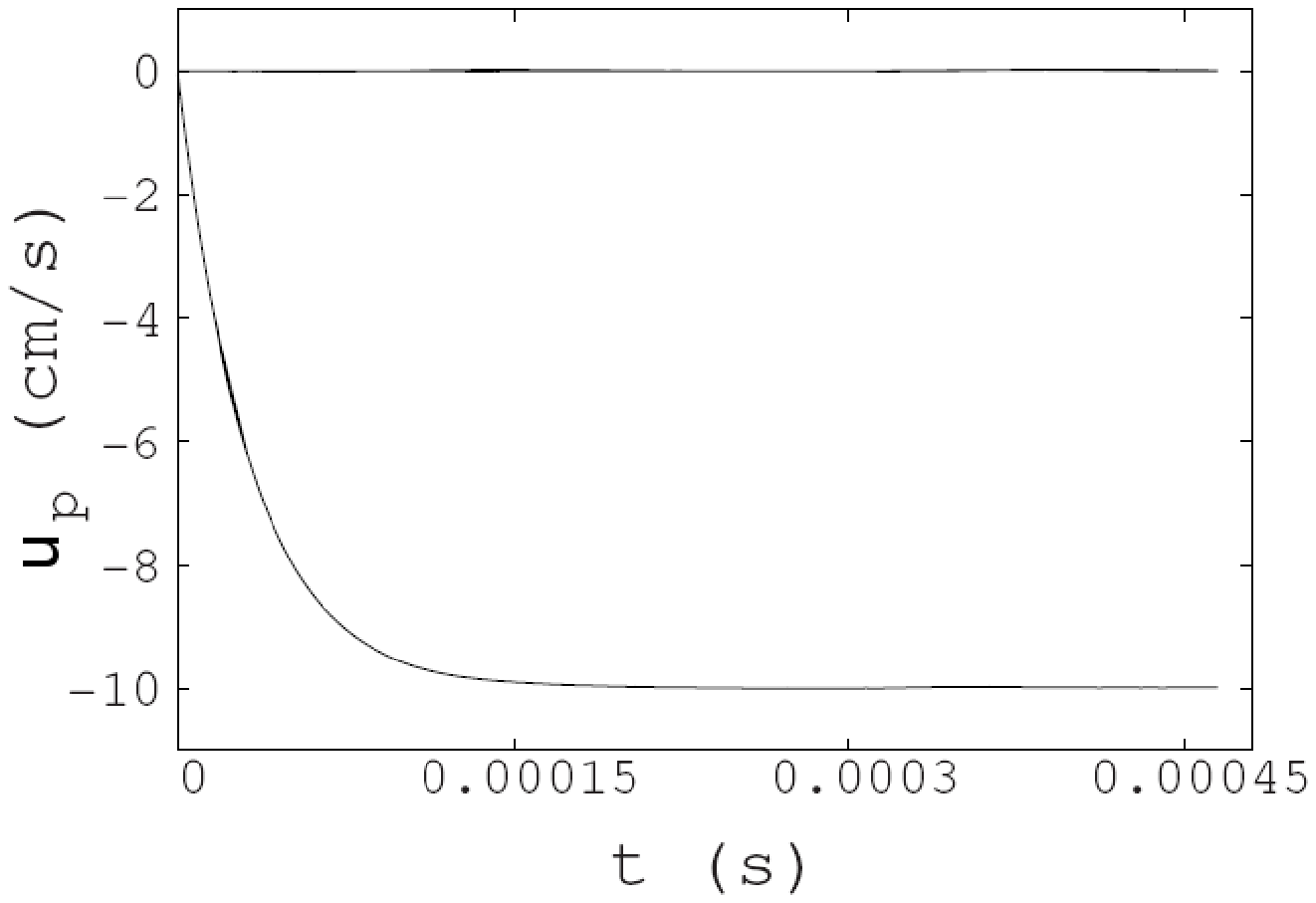}&
\includegraphics[height=0.28\linewidth]{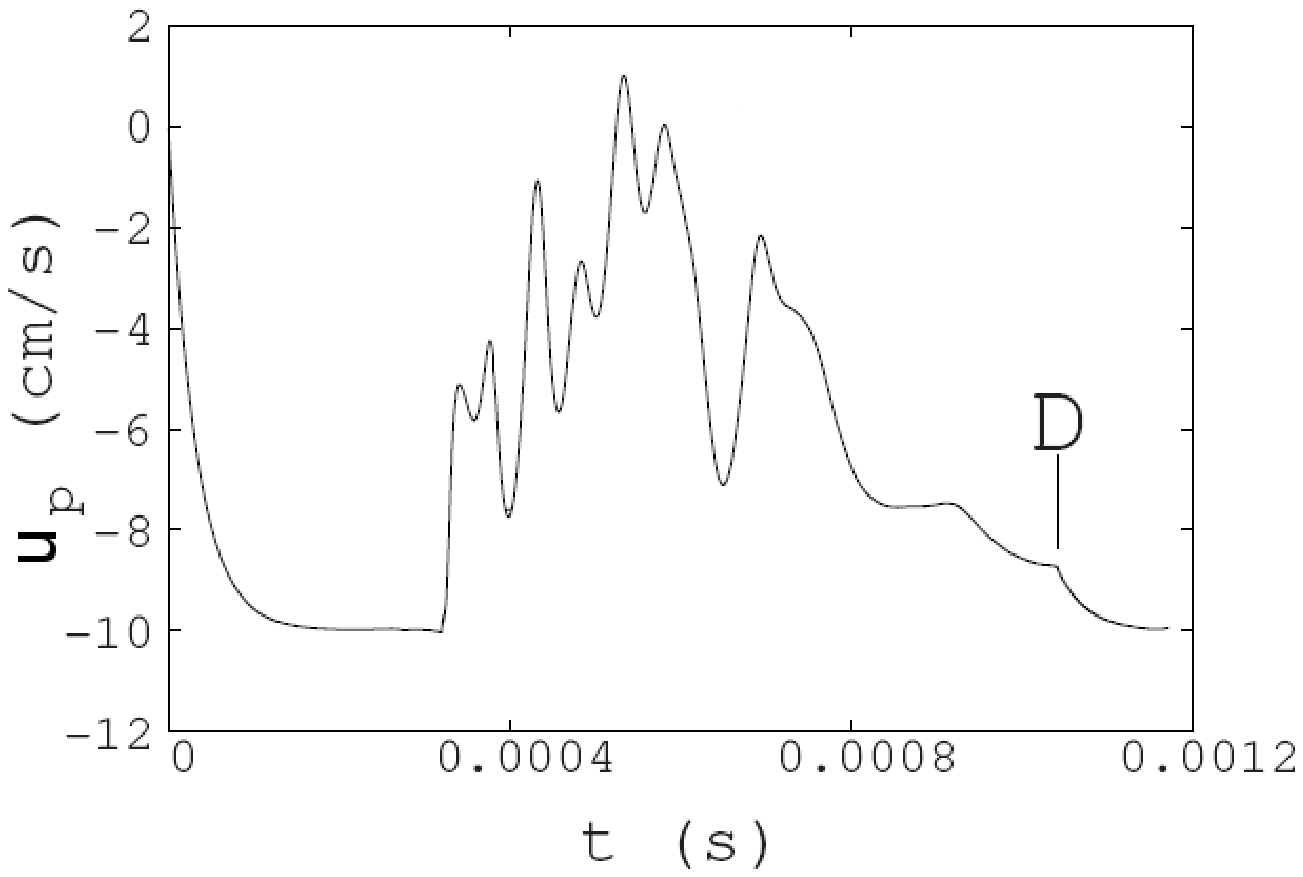}\\
\end{tabular}
\caption{Left: particle velocity  vs time in the absence of particle-vortex
collisions (the transition part of the curve corresponds to the particle response
time). Right: $\up$ vs time illustrating the particle-vortex collision; D indicates
the moment when the vortex detaches from the particle surface and the particle
velocity starts adjusting to the normal flow\cite{K08_b}.
Adapted from Kivotides, \emph{Phys. Rev. B}, \textbf{78}, 224501, (2008).
By permission, \copyright2008
American Physical Society.} \label{Kivotides_08_l_r}
\end{figure}
collision can be identified by strong oscillations (Fig.~\ref{Kivotides_08_l_r}
(right), cf.
the left frame showing the velocity of the particle moving through the tangle
without collisions with quantized vortices) which can be filtered out to restore the
normal velocity.

\subsubsection{Particle tracking experiments} \label{tracking}

Another important visualization experiments in thermal counterflow
are those by Paoletti, Fiorito, Sreenivasan, and
Lathrop\cite{P08_a}. In contrast with the PIV technique, which
analyzes the local average properties of the particulate flow, the
particle tracking technique investigates individual particle
trajectories. Experiments\cite{P08_a} were performed in the
vertical apparatus with the heater at the bottom. The temperature
and the applied heat flux, $q$ ranged from 1.8 to
$2.15\,\textrm{K}$ and from 13 to $91\,\textrm{mW}/\textrm{cm}^2$,
respectively (cf. the experiment of Zhang and Van
Sciver\cite{Z05_1} with $q$ between 110 and
$1370\,\textrm{mW}/\textrm{cm}^2$). Tracers were solid,
micron-size hydrogen particles of density slightly smaller than
that of liquid helium.

It was found that two distinct types of particle trajectories can
be identified: 1)~smooth trajectories corresponding to particles
moving upward in the direction of the normal flow, and
2)~irregular trajectories of particles moving downward; the latter
trajectories were those of particles trapped on quantized
vortices. Experimental observations\cite{P08_a} 
seem to confirm the
earlier theoretical results~\cite{K08} (see above
Sec.~\ref{mechanism_interaction}) that at sufficiently low
relative velocities between the normal fluid and quantized
vortices the particles should be
trapped more easily by the vortex
tangle; at high relative velocities, although the particles
interact with the tangle, it is unlikely that they 
become permanently trapped
on the vortex lines.

\begin{figure}
\begin{center}
\includegraphics[%
  width=0.65\linewidth,
  keepaspectratio]{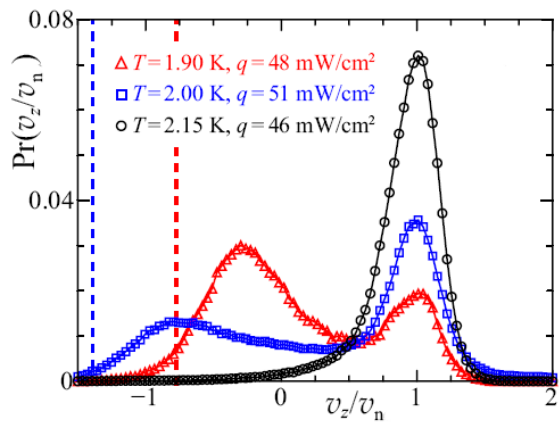}
\end{center}
\caption{(Color online) PDFs\cite{P08_a} of the particle velocity component in
the direction of counterflow. After Paoletti, Fiorito, Sreenivasan, and Lathrop,
\emph{J. Phys. Soc. Japan}, \textbf{77}, 111007, (2008).
Reprinted by permission,
\copyright2008 Physical Society of Japan.} \label{Paol_PDF}
\end{figure}

Calculated from the experimental data\cite{P08_a}, the probability
distribution function (PDF) of the velocity component in the
direction of the counterflow is bimodal, see Fig.~\ref{Paol_PDF},
with the right peak at the normal fluid velocity, and the left,
much broader peak corresponding to particles trapped by the vortex
tangle. Note that, as would be expected, the fraction of particles
trapped by the tangle and hence bimodality of the PDF disappear
either with increasing temperature at constant heat flux (so that
the vortex line density becomes smaller), or with increasing heat
flux at constant temperature (so that $\vn$ becomes sufficiently
high to prevent particle trapping on quantized vortices).

Discussing these experimental findings, Paoletti \emph{et
al.}\cite{P08_a} claimed that their results do not agree with
observations of Zhang and Van Sciver\cite{Z05_1} who found that
the particle velocity is proportional to the normal velocity,
$\up\approx0.5\vn$ independently of temperature. The
authors\cite{P08_a} attributed this discrepancy to Zhang and Van
Sciver's PIV technique which measures the local average properties
of the particulate flow, while the technique employed by Paoletti
\emph{et al.} tracks the motion of individual particles. Paoletti
\emph{et al.} also stressed that their interpretation of
experimental results differs significantly from the theoretical
explanation of Sergeev, Barenghi, and Kivotides\cite{S06_b} (see
also Sec.~\ref{phenomenology} above) whose underlying assumption
was that every particle is affected by quantized vortices as it
moves through the tangle, while the experimental
observations\cite{P08_a} showed that there is a significant
fraction of particles which move freely through the tangle without
experiencing particle-vortex collisions. Furthermore, Paoletti
\emph{et al.} observed a significant temperature dependence of the
particle motion, while the experiment\cite{Z05_1} of Zhang and Van
Sciver and the phenomenological theory\cite{S06_b} of Sergeev,
Barenghi, and Kivotides both show that the disparity between the
particle velocity and $\vn$ is practically independent of
temperature.

Below we will argue 
that the contradiction between the experimental
results of Zhang and Van Sciver\cite{Z05_1} and Paoletti \emph{et
al.}\cite{P08_a}, as well as between theoretical interpretations
of these results by Sergeev \emph{et al.} and Paoletti \emph{et
al.} is only apparent. In fact, it seems that
two recently published works by
Kivotides~\cite{K08_a,K08_b}, reviewed in
Sec.~\ref{self_consistent_counterflow}, already resolve this
apparent contradiction. In these works, based on the two-way
coupling, self-consistent (practically ``first-principle'')
approach, Kivotides has shown that there exist two distinct
regimes of particle motion, one corresponding to a relatively
dense, and another to a relatively dilute vortex tangle. In the
first of these publications Kivotides found that in the case where
the vortex tangle is sufficiently dense
($L=3.284\times10^7\,\textrm{cm}^{-2}$ corresponding to the
intervortex spacing $\ell\approx1.7\,\mu\textrm{m}$ in the
considered example\cite{K08_a}) the average particle velocity is
$0.6\vn$ in a very good agreement with the experimental results of
Zhang and Van Sciver\cite{Z05_1}. Kivotides also showed that in
the considered case the particle is permanently affected by
quantized vortices as it moves through the tangle, and that the
mechanism of particle-vortex interactions agrees with
that suggested by the phenomenological model of Sergeev, Barenghi,
and Kivotides\cite{S06_b}. On the other hand, in the second of his
publications\cite{K08_b} Kivotides, having analyzed numerically a
particle motion in a more dilute tangle
($L=1.168\times10^6\,\textrm{cm}^{-2}$,
$\ell\approx9\,\mu\textrm{m}$), found that in more than half
realizations the particle moves through the tangle with the normal
fluid without interacting with vortices. In the remaining less
than 50\% realizations he observed strong particle-vortex
interactions which in most cases can be described as
trapping-untrapping events. Were it calculated based on the
results reported in the second of his publications\cite{K08_b},
the PDF would have the same bimodal shape as found experimentally
by Paoletti \emph{et al.}\cite{P08_a}.

Although, in terms of the heat flux, the regimes of counterflow in
two reviewed experiments were adjacent, in most observations of
Zhang and Van Sciver the heat flux was an order of magnitude or
more higher than that in the experiments of Paoletti \emph{et al.}
This means that, at the same temperature (e.g. $1.95\,\textrm{K}$
in both experiments), the intervortex spacing in the experiments
of Zhang and Van Sciver ($\ell\approx6\,\mu\textrm{m}$ with
$1.7\,\mu\textrm{m}$ particle in a typical experiment) was at
least an order of magnitude smaller than in the reviewed
experiment of Paoletti \emph{et al.} 
Therefore it can be argued
that two reviewed experimental observations do not contradict each
other but simply correspond to two distinct regimes of particle
motion. The parameter defining each of these regimes is
the ratio of the particle size to the intervortex
spacing, $\ap/\ell$.

As far as the issue of temperature independence of the factor $k$
in the relation $\langle\up\rangle\approx k\vn$ is concerned,
Zhang and Van Sciver's results\cite{Z05_1} are not truly
temperature independent: a closer inspection reveals a relatively
weak dependence of $k$ on temperature (the value $k\approx0.5$ was
obtained by averaging of a large number of experimental data).
Likewise, in the phenomenological theory developed by Sergeev,
Barenghi, and Kivotides\cite{S06_b} this factor is only approximately temperature
independent being a function of the mutual friction coefficient $\gamma$ and of the
parameter $\beta$ which, characterizing the geometry of interactions between the
tangle and the particle, is itself temperature-dependent.

\subsubsection{PIV experiment in thermal counterflow with cylindrical obstacle}
\label{cylinder}

\begin{figure}
\begin{center}
\includegraphics[%
  width=0.65\linewidth,
  keepaspectratio]{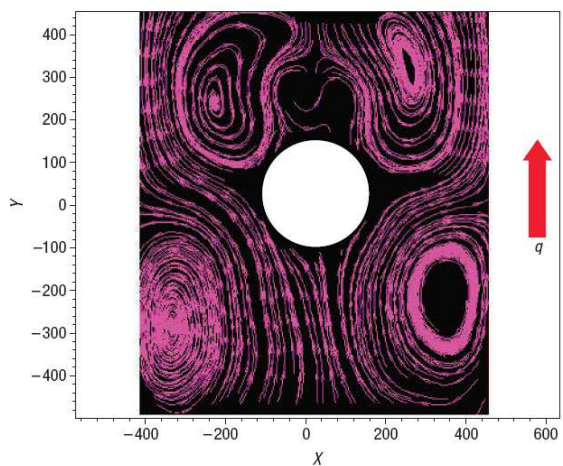}
\end{center}
\caption{(Color online) Streamlines of the particulate motion in the counterflow
around the cylinder\cite{Z05_2} ($q=1.12\,\textrm{W}/\textrm{cm}^2$,
$T=2.03\,\textrm{K}$). From Zhang and Van Sciver, \emph{Nature Physics}, \textbf{1},
36, (2005). Reprinted by permission, \copyright2005 Macmillan Publishers
Ltd.} \label{Van_Sciver_cylinder}
\end{figure}

Among surprising experimental results obtained by the PIV technique
is the recent observation by Zhang and Van Sciver\cite{Z05_2} of the apparently
stationary normal fluid eddies in the thermal counterflow past a
cylinder. In the cited work, Zhang and Van Sciver visualized the
motion of small particles in the thermal counterflow around the
cylinder of diameter $D=0.635\,{\rm cm}$ fixed in the center of
rectangular channel of a cross-section $3.89\times1.95\,{\rm
cm}^2$. The counterflow was produced, in two separate experiments,
by the heat flux $q=0.4$ and $1.12\,{\rm W/cm^2}$ at temperatures
$T=1.6$ and $2.03\,{\rm K}$, respectively (corresponding to the
Reynolds numbers $\textrm{Re}=\rho D\vn/\mun=4.1\times10^4$ and
$2.1\times10^4$). Solid particles used for
visualization in the PIV experiments were polymer microspheres of
diameter $1.7\,\mu{\rm m}$ and density $1.1\,{\rm g/cm^3}$.
In these experiments Zhang and Van Sciver observed the formation of
large-scale eddies of the particulate motion located both downstream
and, surprisingly, upstream of the cylinder with respect to the
normal flow. These, apparently stable
vortices of the particulate flow field were located at distances
about 3 cylinder radii from its center at the angles $\pm45^o$ and
$\pm135^o$ to the axis along the undisturbed flow through the center of the cylinder,
see Fig.~\ref{Van_Sciver_cylinder}. Note that the observed flow structures do not
have a
classical analogue.

Zhang and Van Sciver
attributed the existence of apparently stationary normal eddies
to the mutual friction interaction between quantized vortices and
the normal fluid.
However, our recent 
study\cite{S09} showed that perhaps the experimental
results~\cite{Z05_2} can be interpreted without invoking
the mechanism of interaction between the normal fluid and
quantized vortices. Indeed, the calculation of motion of point vortices in the imposed
potential flow around the circular disk shows that there exist stationary
locations of point vortices, both at the rear and at the front of the
disk. These locations are unstable: any perturbation of the
initial stationary
positions of point vortices leads, eventually, to sweeping of
point
vortices away from their initial locations. Furthermore, some of these stationary
locations are positioned practically as the eddies seen by Zhang and Van Sciver. The
point vortices in the vicinity of such positions will remain close to their initial
locations during the time period corresponding to the duration of the
experiment\cite{Z05_2} and hence seen as apparently stable.

Although this, purely classical explanation of the apparent stability of vortex
structures does not invoke any interaction between the normal and the superfluid
components of $^4$He, the emergence of eddies seen in the experiment\cite{Z05_2}
might still require an explanation based on the analysis of mutual friction between
quantized vortices and the normal fluid.

\section{$^4$He channel flow and turbulent boundary layer} \label{b_layer}

\begin{figure}[t]
\begin{tabular}[b]{cc}
\includegraphics[height=0.35\linewidth]{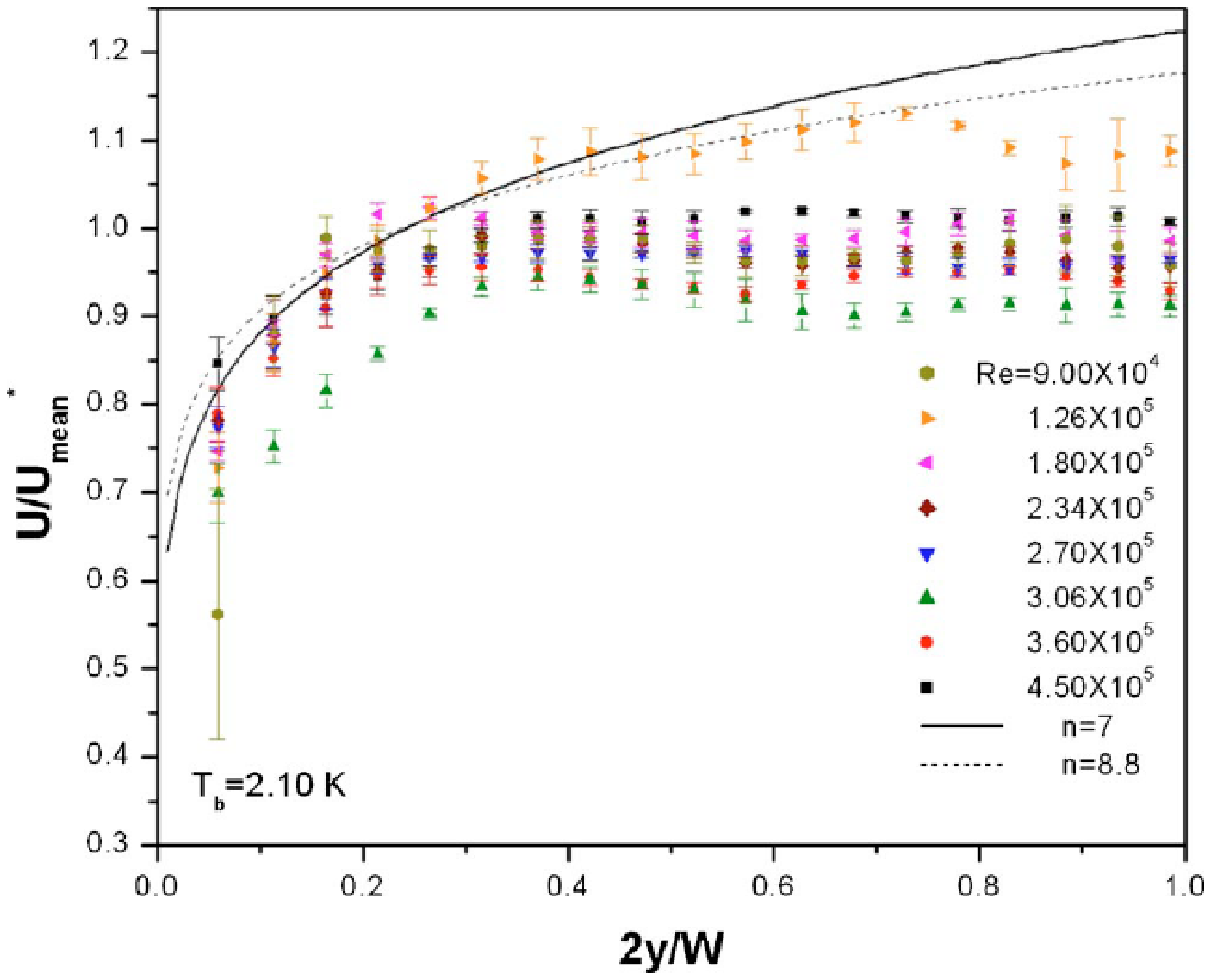}&
\includegraphics[height=0.35\linewidth]{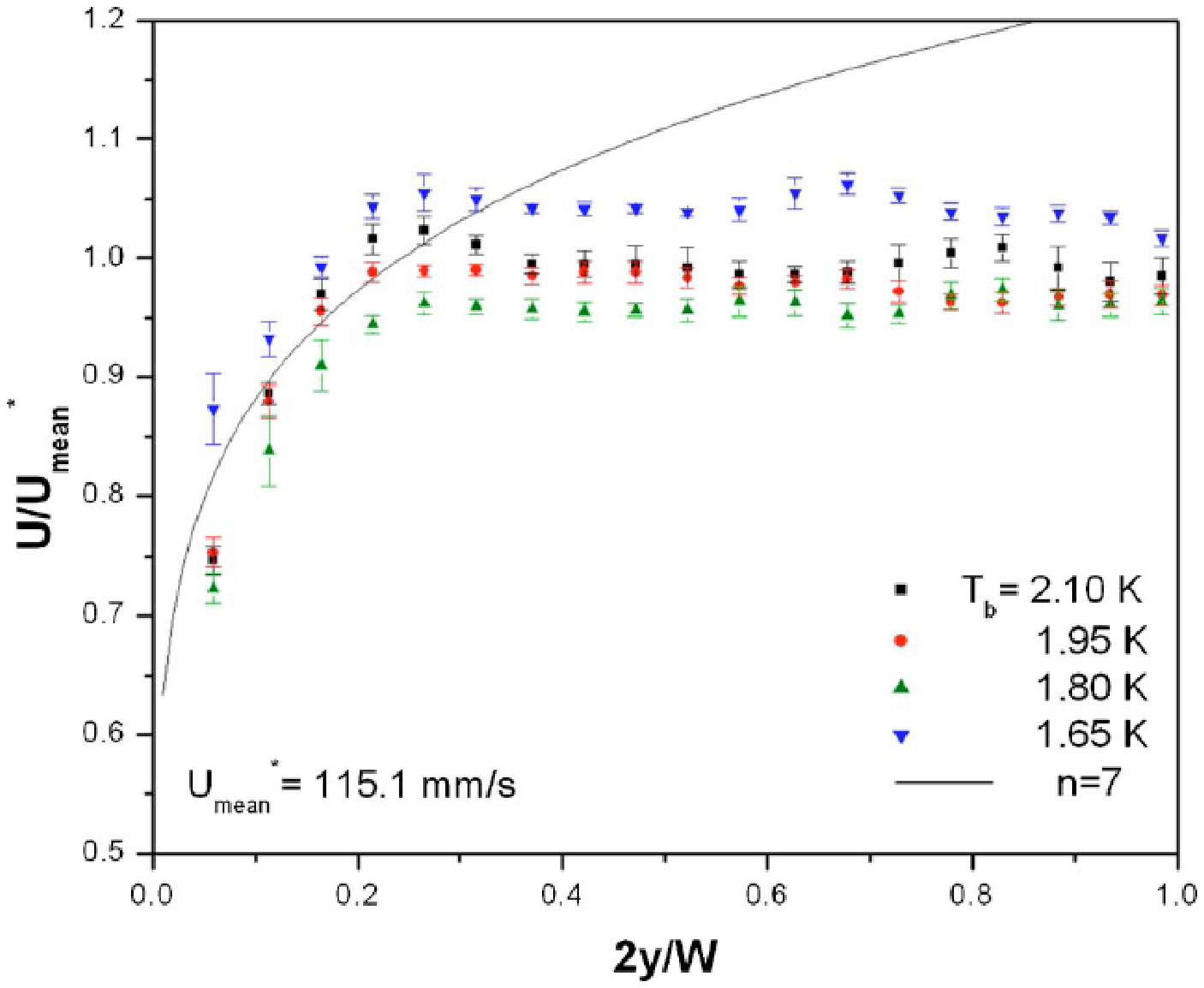}\\
\end{tabular}
\caption{(Color online) Normalized velocity profile\cite{Xu_08} for different
Reynolds numbers at $T=2.10\,\textrm{K}$ (left), and at different temperatures for
the mean velocity $U_{mean}*=11.5\,\textrm{cm}/\textrm{s}$; $W$ is the width of the
square channel. From Xu and Van Sciver, \emph{Physics of Fluids}, \textbf{19},
071703, (2007). Reprinted by permission, \copyright2007 American Institute of
Physics.} \label{boundary_layer}
\end{figure}

This short Section describes the recent experiment which might be a beginning of
systematic study of nonuniform $^4$He flows. Xu and Van Sciver\cite{Xu_08} reported
the PIV measurements, using micron-size deuterium particles, of the $^4$He forced
flow in the rectangular channel for the normal fluid Reynolds numbers ranging from
$9\times10^4$ to $4.5\times10^5$ and temperatures from 1.65 to $2.10\,\textrm{K}$.
In this experiment, at scales larger than the intervortex spacing, the normal and
superfluid components of $^4$He can be considered as fully interlocked so that the
measurements of the particulate velocity field
yield an unambiguous velocity profile of the fluid. The results of Xu and Van
Sciver's measurements were summarized in two graphs\cite{Xu_08} shown in
Fig.~\ref{boundary_layer}.

Xu and Van Sciver concluded\cite{Xu_08} that in the wall region the velocity
distribution agrees reasonably well with the classical $n$th-power law (for $n$
ranging from 7 to 8.8). They also addressed the following questions which yet to be
answered. Why the velocity profile is wider and flatter than that in the classical
viscous channel flow? What is the nature of dependence on Reynolds number of the
results shown in Fig.~\ref{boundary_layer} (left)? Why, as seen from
Fig.~\ref{boundary_layer} (right), the normal fluid density does not seem to affect
the shape of velocity profiles?

\section{Visualization of vortex reconnections. Velocity statistics in decaying
quantum turbulence} \label{reconnections}

\begin{figure}
\begin{center}
\includegraphics[%
  width=0.95\linewidth,
  keepaspectratio]{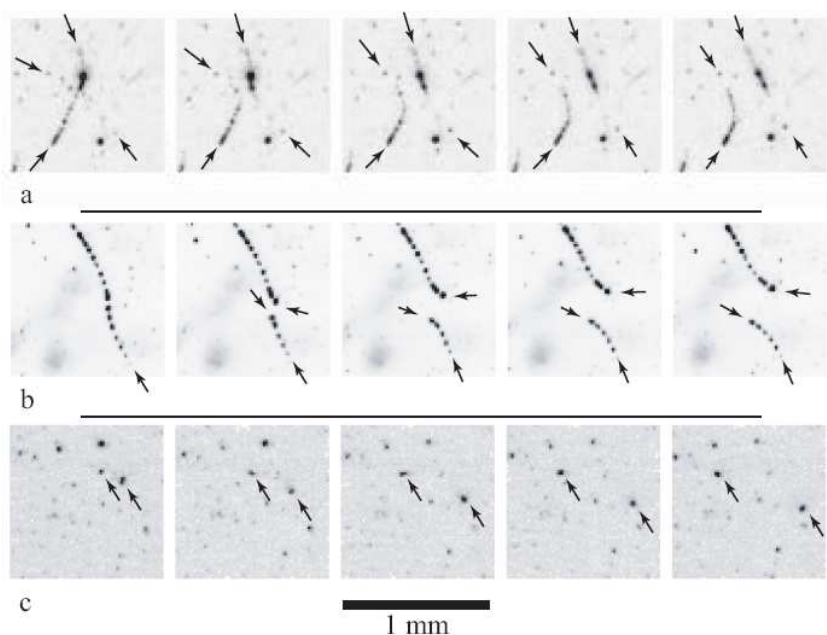}
\end{center}
\caption{(Color online) After Bewley, Paoletti, Sreenivasan, and Lathrop
(\emph{Proc. Nat. Acad. Sci.}, \textbf{105}, 13707, (2008)):
visualization\cite{B09_b} of the vortex dynamics. The
particles trapped on quantized vortices can be easily identified. In each of the
sequences, a, b, and c, the images were taken at $50\,\textrm{ms}$ intervals.
(a):~two approaching vortices have several particles trapped on their filaments (the
first frame shows the projection in which the vortices appear crossed). The
sequence~(b) seems to show the vortices moving apart after the reconnection. The
sequence~(c) illustrate the authors'\cite{B09_b} method of identification of
reconnecting vortices by a sudden motion of two tracer particles away from each
other. \copyright2008 USA National Academy of Sciences.}
\label{recon_images}
\end{figure}

In a recent experiment\cite{B09_b} Bewley, Paoletti, Sreenivasan, and Lathrop
observed the motion of solid hydrogen particles trapped on quantized vortices. The
specific purpose of this work was a direct experimental investigation of vortex
reconnections in turbulent $^4$He. A sequence of images\cite{B09_b} illustrating the
motion of particles trapped on vortex filaments is reproduced in
Fig.~\ref{recon_images}.

To quantify their results, Bewley \emph{et al.}\cite{B09_b} assumed that the
evolution of reconnecting vortices can be characterized by a single scale parameter,
$l(t)$. Using, as a measure of $l(t)$, the experimentally observed distance between
two particles closest to the point of reconnection, they found that the evolution
obeys the scaling $l\sim(t-t_0)^{1/2}$, where $t_0$ corresponds to the moment of
reconnection of two vortices.

\begin{figure}
\begin{center}
\includegraphics[%
  width=0.65\linewidth,
  keepaspectratio]{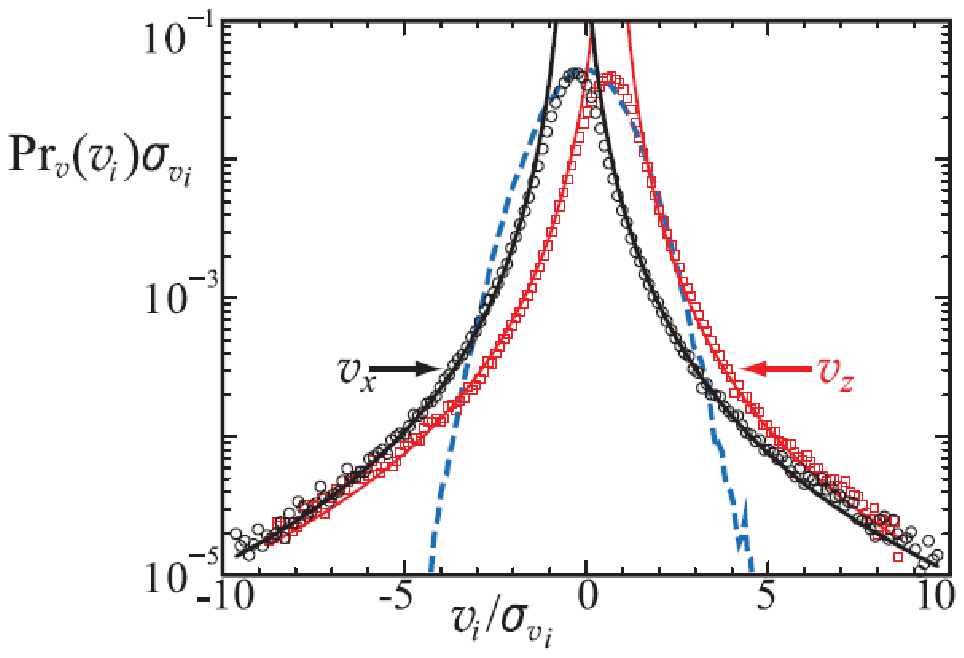}
\end{center}
\caption{(Color online) Probability distribution functions\cite{P08_b} of the velocity
components in the direction of ($v_z$) and normal to ($v_x$) the counterflow. The
distributions are scaled with $\sigma_{v_z}=0.074\,\textrm{cm}/\textrm{s}$ and
$\sigma_{v_x}=0.066\,\textrm{cm}/\textrm{s}$, respectively. The dashed (blue) line
-- velocity PDF, scaled with $\sigma_{v}=0.25\,\textrm{cm}/\textrm{s}$, for
classical turbulence. From Paoletti, Fisher, Sreenivasan, and Lathrop,
\emph{Phys. Rev. Lett.}, \textbf{101}, 154501, (2008).
Reprinted by permission, \copyright2008 American Physical
Society.} \label{vel_stat}
\end{figure}

The particle tracking technique was further developed by Paoletti, Fisher, Sreenivasan, and
Lathrop~\cite{P08_b} to investigate, by analyzing the trajectories of tracer
particles, the velocity statistics in decaying quantum turbulence. The decay of
turbulence produced initially by the thermal counterflow was studied after the
counterflow has been stopped by switching the heater off. Paoletti \emph{et al.}
found that the PDF of the particle velocity is strongly non-Gaussian with a
pronounced tail obeying $v^{-3}$ power law, see Fig.~\ref{vel_stat}.
The authors attributed such a tail to high velocities 
produced by reconnections of quantized vortices.
The experiment has stimulated
calculations \cite{Angela} of turbulent velocity statistics in
the context of three-dimensional and two-dimensional atomic 
Bose-Einstein condensates; after finding
similar non-Gaussian velocity statistics in these related quantum systems
the authors argued that, in general, non-Gaussian statistics arise 
from the singular nature
of the velocity field around quantized vortices.

The results of fundamental importance reported in these two reviewed
publications~\cite{B09_b,P08_b} raise, however, the following question: what effect
have the particles trapped on vortex filaments on the evolution of quantized
vortices and, in particular, on the reconnection of two approaching vortices? In
their work\cite{P08_a} reviewed earlier in Sec.~\ref{tracking} the authors claimed
that, because in their experiments the distance between micron-size particles
trapped on the same vortex filament was typically about $100\,\mu\textrm{m}$, the
influence of trapped particles on the evolution of vortices can be neglected. This
claim is yet to be justified
(especially where the motion and evolution of vortex filaments near the reconnection
point is concerned). It is unlikely that at present this problem can be treated
experimentally; a more realistic alternative seems to be a development of the
mathematical model of a vortex filament
loaded by trapped solid particles\cite{ring_particles}.

\section{Related techniques: Detection of vortices by trapping of
negative ions and imaging of $\textrm{He}_2$ molecules}
\label{related_techniques}

\subsection{Ion trapping} \label{ion_trapping}

The technique of vortex detection by ion trapping was recently
reviewed in a number of regular research publications including some
in this
journal\cite{Golov-experiments,Golov-charged_rings},
so that we will present here a rather brief overview of theoretical
and experimental aspects of this technique.

The idea of using charge carriers for detection of quantized
vortices in superfluids is much older than that of using solid
particles for the flow visualization in $^4$He and dates back to
the works of Careri\cite{Careri}, and Reif and
Meyer\cite{Reif_Meyer}. The technique is based on the phenomenon
that a negative charge (a single electron) injected into $^4$He
self-localizes itself in a spherical void (known as the ion or
electron bubble) of radius $12-20\,\textrm{\AA}$ (depending on
pressure) from which helium atoms are excluded. (A relatively
large size of this, almost macroscopic bubble justifies inclusion
of this Chapter into our review otherwise concerned with the
motion of much larger, solid particles in $^4$He.) The model of
the negative ion bubble was proposed by Ferrel and later
elaborated by Kuper and other authors\cite{bubble_model} and
confirmed experimentally by Levine and
Sanders\cite{Levine-Sanders}. Consistent with the bubble model of
the negative charge carrier is the effective mass of the ion,
$M^*$ determined in Ref.~\cite{effective_mass}. It appears in the
Langevin equation of motion for the ion's drift velocity, $\uu_D$
in the electric field, ${\bf E}$: $e{\bf
E}=M^*\left(d\uu_D/dt+\uu_D/\tau\right)$, where $e$ is the
electron charge, and $\tau$ the phenomenological relaxation time.
The effective mass of the ion was found\cite{effective_mass} to be
of the order 100 masses of $^4$He atom (in agreement with the
bubble model), and $\tau$ appears to be the viscous relaxation
time introduced above in Sec.~\ref{Lagrangian_equations}. (The
detailed treatment for the Stokes problem of the ion bubble
motion, albeit in normal $^4$He, was given by Ostermeier and
Schwarz\cite{Ostermeier_Schwarz}.)

The possibility of using negative ions as detectors (or ``probe
particles'') of quantized vortices is due to the fact that the ion
bubble is attracted to the vortex by the Bernoulli force. The combined
action of the Bernoulli effect and the reduced condensation energy of
the core produces a potential well of the depth of the order
$50\,\textrm{K}$\cite{Donnelly-books} and hence traps the ion.
This mechanism\cite{Donnelly-books} is practically identical to that
described above in Sec.~\ref{trapping} of trapping of solid particles
on the vortex cores. (Note that the mechanism of trapping of the ion
bubble by the quantized vortex in $^4$He modeled as the Bose
condensate was studied by Berloff and Roberts\cite{Be00} and described
above in Sec.~\ref{evidence}.)

The model of the positive charge carrier in $^4$He was developed
by Atkins\cite{Atkins} in 1959 and became since then commonly
accepted. According to this model the positive ion exerts an
electrostrictive attraction on the surrounding fluid thus causing
a liquid-solid transition resulting in a core (``cluster'' or
``snowball'') of solid helium. The radius of this snowball is from
7 to $9\,\textrm{\AA}$, depending on pressure. Like the negative
ion bubble, such a snowball will be attracted, by the Bernoulli
force, to the vortex core so that, in principle, positive ions can
also be used for detection of quantized vortices. However, a
smaller than that of the ion bubble radius of the snowball and,
therefore, the cross-section of the snowball-vortex interaction
make the experimental technique using positive ions more difficult
and less practical. Hence in this review we will discuss only
works concerned with negative charge carriers.

The realization of the negative ion technique can be illustrated by
the following arrangements of the experiment, typical of the early
observations of single vortex lines in rotating $^4$He: quantized
vortex lines are charged first by ion bubbles trapped on them using
e.g. the electron beam emitted orthogonally to the axis of rotation.
The vortex lines are then detected by applying the electric field
parallel to the axis of rotation so that trapped electron bubbles
slide along vortex filaments to a collector attached to an
electrometer. The latter registers the amount of collected charge
which is proportional to the number of vortex lines present. In its
early, simplest version this technique allowed to determine only the
total number of (almost) straight quantized vortex lines in a slowly
rotating container. Its subsequent modifications made possible also
detection of individual vortex lines and later the measurement of the
vortex line density in turbulent $^4$He.

The theory of ion trapping by the quantized vortex has been
developed in classical works of Donnelly, Roberts, and
Parks\cite{Parks-Donnelly,Donnelly}. To analyze ion-vortex
interactions, two competing mechanisms were taken into account:
trapping of the ion bubble, moving in the electric field, in the
potential well of the vortex, and escape of the trapped ion due to
its Brownian motion, which was considered to be in equilibrium
with thermal (quasiparticle) excitations in $^4$He. Two parts of
this problem were calculations of 1)~the trapping (or capture)
cross-section, $\sigma,\,\textrm{cm}$, and 2)~the escape
probability, $P,\,\textrm{s}^{-1}$; the second part was analyzed
using Smoluchowski or Fokker-Planck equations in the framework of
the Kramers-Chandrasekhar method for calculation the probability
of escape of a particle by diffusion from a potential well. The
capture cross-section, which was found to be of the order
$10^{-6}$ to $10^{-5}\,\textrm{cm}$ and to decrease with the
electric field and increase with temperature to about
$1.6\,\textrm{K}$, at which temperature it drops sharply. The
escape probability (or, equivalently, the mean trapping lifetime,
$t_\ell\sim P^{-1}$) was found in a good agreement with
experimental data obtained by Springett, Tanner, and
Donnelly\cite{Springett,Tanner} as well as with experimental
results of Douglass and Cade\cite{Douglass-Cade}. The effective
cross-section can then be calculated as $\sigma_{eff}=\sigma
e^{-Pt}$, where $t$ is some characteristic time. Parks and
Donnelly\cite{Parks-Donnelly} further developed the theory in
order to make possible calculation of the ion bubble radius from
the experimentally measured mean trapping lifetime. This required,
in particular, a calculation of the so-called substitution energy,
i.e. a kinetic energy of rotating superfluid excluded by the
trapped bubble. For the ion bubble of radius $R$ trapped
symmetrically on the vortex core this was found as
\begin{equation}
\Delta E=\frac{\rhos\kappa^2R}{2\pi}
\left[1-\left(1+\frac{\xi^2}{R^2}\right)^{1/2}
\sinh^{-1}\left(\frac{R}{\xi}\right)\right] \label{substitution}
\end{equation}
(note that formula~(\ref{reduced_E}) of
Sec.~\ref{trapping_mechanism} follows from
Eq.~(\ref{substitution}) assuming $R=\ap\gg\xi$). In the context
of further experimental studies discussed below, it is worth
noticing that the substitution energy (equal to the depth of the
potential well) is about $50\,\textrm{K}$. Parks and Donnelly's
study of bubble radii was further developed by Springett and
Donnelly\cite{Donnelly-pressure} who used the measurements of
trapping cross-sections in the rotating container to deduce that
the radius of the ion bubble decreases with pressure. The theory
developed by Donnelly \emph{et al.} also predicted that the mean
trapping lifetime should increase with the superfluid density and
with the radius of ion bubble. These predictions were soon
confirmed experimentally by Springett\cite{Springett} who
measured, at various pressures and temperatures, the cross-section
of the ion capture and ion mobilities in order to derive from
these data the pressure and temperature dependence of the ion
bubble radius and the mean trapping lifetime. Later Pratt and
Zimmermann\cite{Pratt} also measured the mean trapping lifetime in
the wide range of temperatures and pressures (from vaporization
and solidification) and showed that at constant pressure the
lifetime rapidly decreases with $T$. In particular, it was found
that at saturated vapor pressure trapping becomes negligible at
temperature above $1.7\,\textrm{K}$, the value which is now often
referred to as the ``abrupt lifetime edge''. It was also shown
that the temperature, below which trapping becomes significant
increases with pressure. Glaberson\cite{Glaberson} measured the
mobility, as a function of temperature and pressure, of negative
ions trapped on quantized vortex lines, and arrived at the
important conclusion that the negative ion bubbles do not deform
as they are trapped on vortex cores. He also developed a new
model, which appeared to account satisfactory for available
experimental results, for the drag exerted on the trapped ion
bubble, based on the assumption that the vortex line has  several
\AA thick central core surrounded by a tail of excess roton
density with momenta opposite to the direction of circulation.

Among the first applications of the ion trapping technique for an
investigation of vortex structures in $^4$He was an observation by
Northby and Donnelly\cite{Northby} of a nonlinear dependence of
the number of quantized vortex lines on the angular velocity of
the rotating container. This nonlinearity was found to be
associated with the existence of a near-wall region, observed in
the cited work, free of quantized vortices.

The early experiments\cite{Careri,Vicenti-Misoni} have indicated that
ions interact strongly with turbulence in $^4$He, but no quantitative
data have been obtained yet. The first experimental study of the
turbulent vortex tangle by the ion trapping technique was undertaken
by Sitton and Moss\cite{Sitton-Moss} in 1969 (in fact it was the first
direct experimental confirmation that the turbulence in the superfluid
component of $^4$He is composed of individual, quantized vortex lines
indistinguishable from those produced in the (slowly) rotating
container except for their configurations). The vortex tangle was
produced by the supercritical heat current at temperature above
$1.6\,\textrm{K}$. Since most of the vortex lines are no longer
straight but have a configuration of loops and kinks, the charge can
no longer slide along the filaments to a collector but is trapped
inside the tangle. The fraction $f_Q$ of the
trapped charge was measured and
then linked with the vortex line density by the relation
$f_Q=(1+P/u_i\sigma L)^{-1}$, where $u_i$ is the ion velocity, and
$L$ can be linked with the heat flux and properties of $^4$He by the
well known Vinen equation\cite{Vinen-equation}.

A series of experimental studies and their interpretation,
beginning 1972 and spanning the period of nearly 30 years, was
undertaken by the group led by Packard and Williams. In 1972,
Packard and Sanders\cite{Packard-Sanders} showed the possibility
of detecting individual vortex lines in rotating $^4$He. The
experiments were performed in a cylindrical container whose
rotation slowly accelerated so that the number of vortex lines
present increased with time. The experimental arrangements were
similar to those described above,.i.e. the charge trapped on
vortex lines was measured by applying the axial electric field
thus transferring the charge to a collector attached to an
electrometer. The appearance of each new vortex line was detected
by a steplike increase in the electrometer's reading. Bringing the
container slowly to rest, the authors also discovered the
existence of remanent vorticity studied later by Awschalom and
Schwarz (see below).

Later Williams and Packard\cite{Williams-photo} developed the
first photographic technique which directly visualized spatial
positions of individual quantized vortex lines in rotating $^4$He.
The experimental arrangements were similar to those of
Ref.~\cite{Packard-Sanders}, but in addition the magnetic focusing
was used to stabilize beams of electrons emitted from the charged
vortex lines; these beams impinged on a phosphor screen so that
the positions of quantized vortices could be actually
photographed. Experiments were performed at temperatures lower
than $0.3\,\textrm{K}$, and up to 0.8\% of $^3$He (which acted as
a fixed amount of normal fluid) was added to stabilize the
vortices. Among other results, it was found, in agreement with the
classical Feynman's prediction, that the average intervortex
spacing is $\sim\sqrt{h/2\omega m_4}$, where $\omega$ is the
angular speed of rotation, but in this work an expected stable
triangular lattice of vortices was never observed. The photographs
of regular, symmetric arrays of vortices were obtained when this
experiment was repeated some years later\cite{Yarmchuk}. (These
photographs should be very familiar to a reader through numerous
reproductions in other publications and hence are not shown here.)
The authors ascribed to mechanical disturbances their earlier
failure to observe the regular structures. They also argued that
the symmetric state of the system is determined not by the
absolute minimum of the free energy, as has been believed earlier,
but by the combination of the past history of the system and the
local minimum of the free energy, concluding that the symmetric
state is highly metastable. The detailed description of the
photographic technique for visualizing the positions of quantized
vortices in rotating $^4$He was given in
work\cite{Williams-photo-improved}. Much later one of the authors
claimed\cite{Williams-Zeit} that usual helium ions are not really
suitable for detection of thermally excited vortices, and that
multielectron bubbles should be employed for this purpose. Also
such an experiment was outlined in the cited paper, the authors of
this review are not aware of any practical development in this
direction.

Using the experimental technique of Ref.~\cite{Williams-photo},
this research group also revisited\cite{Williams-DeConde} the
problem, analyzed earlier theoretically by Donnelly, Roberts, and
Parks\cite{Parks-Donnelly,Donnelly}, of the lifetime of ions
trapped on the vortex lines. Theoretical analysis predicted that
in the temperature interval $0.6<T<1.1\,\textrm{K}$ the mean
lifetime should be longer than $10^{13}\,\textrm{s}$ and that it
should increase with decreasing temperature. However, the
experiments revealed that the observed lifetime was many orders of
magnitude smaller and was actually decreasing as temperature
decreased. To resolve the contradiction with the theoretical
predictions, the authors suggested that at temperatures lower than
$1.5\,\textrm{K}$ the trapping lifetime is no longer determined by
the intrinsic properties of bubble-vortex interaction but by a
time scale of the vortex motion until it is destroyed by the
container's wall. The details of this mechanism were suggested by
the authors as follows: as the moving vortex filament encounters
the wall, it is destroyed (so that the charge is collected by the
wall), but, to maintain the equilibrium value of the vortex line
density, another, uncharged filament is created. In the case where
the time scale of migration of the vortex filament to the wall is
smaller than the intrinsic trapping lifetime of the ion bubble,
the vortex motion will be a limiting factor of the charge loss.
The authors concluded that at low temperatures such that the
normal fluid density and hence viscous damping (by the mutual
friction) become negligible, the observed lifetime is actually the
measure of the timescale of the vortex migration to the wall and,
therefore, should be independent of temperature, in agreement with
the authors' observations. Furthermore, the authors argued that
their conclusions can also be related to the earlier observations
by Cheng, Cromar, and Donnelly\cite{Donnelly-Cheng} that the ion
trapping reduces in the presence of the axial heat current. The
explanation suggested in Ref.~\cite{Williams-DeConde} is that in
the counterflow turbulence the mutual friction increases the
intensity of the vortex motion and hence the rate of destruction
of vortices by the wall.

Employing the ion trapping technique, one of the works of key
importance for understanding the mechanism of the onset of quantum
turbulence is that of Awschalom and
Schwarz\cite{Schwarz-remanent}. This experiment used two parallel
plates, immersed in $^4$He, one of which serves as a charge
collector detecting the vortices pinned to both plates. (The
experiment aimed, in particular, to support the earlier Schwarz's
idea\cite{Schwarz-multiplication}, based on his analysys of vortex
reconnections, that under some conditions vortex singularities can
multiply.) The experiment showed that quiescent $^4$He contains a
rather large number of quantized vortex filaments which are pinned
metastably to the parallel plates. The line density of these
remanent vortices seems to be history independent, and its
existence was established upon going, in temperature, down through
the  $\lambda$ transition. (Note that in this work the vortex line
density was not measured directly but estimated, considering the
rather complicated geometry of electric field lines, from the
measurement of the collected charge.) Awschalom and Schwarz found
that the line density of remanent vortices is very close to the
critical line density, introduced by Tough\cite{Tough}, below
which vortices disappear. The authors' interpretation of the
critical line density was that the vortex filament do not
disappear but become immobilized by pinning to the walls and hence
cannot be observed by any conventional experimental technique. The
authors also justified their ideas by analyzing the evolution of
the vortex line density in decaying turbulence produced by
ultrasound. They observed that, after the ultrasound has been
switched off, $L$ decays not to zero but to the value of the
remanent vortex line density. (Earlier experiments of Milliken
Schwarz, and Smith\cite{Schwarz-Milliken-remanent} produced
qualitatively similar results.) The authors concluded that any
volume ``will be penetrated \emph{ab initio} by quantized vortices
stabilized by surface pinning'', and hence all experiments on the
evolution of the vortex line density should be interpreted as if
$L$ starts from the value corresponding to the remanent vortex
line density.

The first attempt of systematic experimental study, by means of
the ion trapping technique, of the decay of quantum turbulence was
made by Davis, Hendry, and McClintock\cite{McClintock-decay}. In
turbulence, generated by the oscillating grid, some ions emitted
from the tip get trapped on vortices within the tangle and hence
reduce the current arriving at the collector. The evolution of the
vortex line density can be estimated from the evolution of this
current, and the experiment clearly demonstrated the production
and decay of quantum turbulence as well as its spatial
distribution. However, this experiment had certain disadvantages.
In particular, the value of the vortex line density in the low
temperature limit (from 200 down to $22\,\textrm{mK}$) remained
unknown, and the data on trapping cross-sections at these
temperature had not yet been available. Nevertheless, thies
experiment yielded a very useful information on the time scale of
decay, and also showed that in the temperature range from 22 to
$70\,\textrm{mK}$ the process becomes temperature independent.

Clearly, a modification of the classical ion trapping experimental
arrangements was required that would make possible, in the low
temperature limit $T<200\,\textrm{mK}$, to measure directly the
dynamics of the vortex tangle as well as the trapping
cross-sections. Such a new technique was recently developed by
Walmsley, Golov and their
co-workers\cite{Golov-experiments,Golov-charged_rings,Golov-technique,Golov-decay};
the detailed description of the experimental cell and methods can
be found in the first two of the cited works.

The experimental cell, mounted on a rotating cryostat, is a cube
whose sides are electrodes serving as the charge collectors. In
the earlier experiment\cite{Golov-experiments} just one ion
emission tip was fitted at the bottom side of the cell, while
later
experiment\cite{Golov-charged_rings,Golov-technique,Golov-decay,Golov-ultraquantum}
used two emission tips fitted at the bottom and one of the side
plates. In order to provide the means of measuring the spatial
properties of the vortex tangle, a difference of electric
potential was kept between some of the electrodes to ensure
depletion of the ion current emitted by the tips. An analysis of
ion trapping by quantized vortices suggests the exponential decay
of the ion current, collected by the electrodes, with the vortex
line density, so that the latter can be recovered from the
measurements of the current $I(t)$ from the relation
$L(t)/L_0=(\sigma d)^{-1}\ln[I(\infty)/I(t)]$, where $d$ is the
size of the experimental cell. The turbulence was produced by spin
up or/and spin down of the rotating experimental cell.

In Ref.~\cite{Golov-experiments} the authors argued in favor of
using, as probe particles, the charged vortex rings rather than
bare ions. The reason is that the trapping cross-section of the
bare ion bubble is small,
$\sigma\sim10^{-6}\,\textrm{cm}$\cite{Donnelly,Glaberson-cross-section},
and was never measured at temperature below $0.8\,\textrm{K}$. On
the other hand, at low temperatures, in the case where the vortex
nucleation velocity, $v_c$ is lower than Landau critical velocity,
$v_L\sim5\times10^3\,\textrm{cm/s}$, the ion, upon reaching the
velocity $v_c$ will nucleate the vortex ring that captures the ion
thus forming the stable ion-ring complex. The details of this
mechanism can be found in the classical paper by Rayfield and
Reif\cite{Rayfield-Reif} (see also the work by Nancolas, Bowley,
and McClintock\cite{McClintock-charged_rings} and Donnely's
monographs\cite{Donnelly-books}). While attempting, initially, to
repeat the experimental conditions of the work by Davis, Hendry,
and McClintock\cite{McClintock-decay}, who tried to eliminate
nucleation of vortex rings by ions, Walmsley, Golov and their
co-authors eventually came to a conclusion that at low
temperatures charged rings are preferable as proble particles. The
trapping cross-section of the charged ring is of the order of its
diameter,
$\sigma\sim1\,\mu\textrm{m}$\cite{Scwarz-ring_cross-section},
which is much larger than the trapping cross-section of the bare
ion bubble, and hence it is considerably easier to detect
quantized vortices by charged rings (as was stated much earlier by
Guenin and Hess\cite{Guenin}). Some disadvantage of using charged
vortex rings is their dynamics which is more complicated than that
of bare ion bubbles (we refer the reader to the theoretical
studies\cite{charged_ring-theory}).

The experimental
studies\cite{Golov-experiments,Golov-charged_rings} of Walmsley
\emph{et al.} showed that using charged rings rather than bare
ions it is possible to detect the change in the vortex line
density upon starting and stopping rotation at temperatures from
0.5 down to $30\,\textrm{mK}$. Monitoring in
Ref.~\cite{Golov-charged_rings} the trapping of charged rings by
rectilinear vortices generated by slow rotation the authors were
able to measure the ring-vortex trapping cross-section at these
temperatures. Furthermore, at low temperatures injection of
charged rings can be used to create a vortex
tangle\cite{Golov-technique}.

An application of this technique to the study of quantum
turbulence in the limit $T\to0$ has already brought some
non-trivial results. Thus, Walmsley \emph{et
al.}\cite{Golov-decay} studied, at temperatures from 1.6 down to
$0.08\,\textrm{K}$, a decay of the homogeneous vortex tangle
produced by a sudden spin down of the rotating cell. They found
that the decay of the energy flux, $\epsilon$ is the same as that
of the classical turbulence at high Reynolds numbers, i.e.
$\epsilon\sim t^{-3}$, and that the vortex line density decays as
$t^{-3/2}$. Most importantly, it was found that at
$T\approx0.8\,\textrm{K}$ the effective kinematic viscosity,
$\nu_K$ drops sharply from the value $\nu_K\approx0.1\kappa$ and
approaches $\nu_K\approx0.003\kappa$ as $T\to0$. The authors
linked this drop to the transition to the new form of turbulence,
the so-called ``ultraquantum'' (or ``Vinen'' as opposed to the
classical, ``Kolmogorov'') turbulence. This, less structured than
(quasi)classical regime of turbulence is characterized by the
cascade in which the energy is transferred to smaller scales by
Kelvin waves on individual vortex lines (and, therefore, by the
absence of any large scale motion). The mechanism of energy
dissipation is though to be acoustic: smallest perturbations are
emitted as phonons. Clearly, the ultraquantum regime should be
characterized by an effective kinematic viscosity, $\nu_V$
different from the quasiclassical $\nu_K$.

It should be emphasized, though, that the regime of turbulence in
the cited work\cite{Golov-decay} was essentially quasiclassical
due to forcing at large scales (rotation). The existence and
properties of ultraquantum regime at temperatures below
$0.5\,\textrm{K}�$ were addressed a year later in the recent work
of Walmsley and Golov\cite{Golov-ultraquantum}. The vortex tangle
with was produced not by spin down of the rotating scale but by
charged vortex rings so that no large-scale flow was generated. It
was found that in the ultraquantum regime the vortex line density
decays as $t^{-1}$, and that the effective kinematic viscosity of
the Vinen turbulence is $\nu_V\approx0.1\kappa$.

In conclusion of this Section we would like to emphasize that the
considered so far two experimental techniques of visualization
(or, at least, detection) of flow properties in turbulent $^4$He,
i.e PIV/particle tracking and the ion trapping technique do not
compete but rather complement each other. Indeed, the ion trapping
technique detects quantized vortices without disturbing the flow,
and it is efficient in the low temperature regime; at temperatures
above $1.7\,\textrm{K}$ ion bubbles can no longer be trapped on
quantized vortices. As strongly emphasized by Charalambous
\emph{et al.}\cite{Charalambous}, the obvious disadvantage of this
technique is that ion bubbles ``cannot provide the kind of
detailed information about turbulent flow patterns that is
available in the case of classical turbulence''. We fully endorse
this statement. Actual visualization of quantized vortex filaments
is also hardly possible (but see some new developments discussed
below). On the contrary, solid, micron-size particles being used
in the PIV and particle tracking techniques, albeit disturbing the
flow, can visualize the normal flow and can even be used to
``paint'' quantized vortices. However, as was demonstrated in
Secs.~\ref{instability} and \ref{Tto0}, in the low temperature
regime, as $T\to0$ solid particles do not follow the flow, and, as
was shown in Sec.~\ref{temperature}, most likely cannot be trapped
on quantized vortices.

In the remainder of this chapter we will briefly discuss two new
promising techniques that potentially can be used for direct flow
visualization.

\subsection{Cavitating electron bubbles and metastable
$\textrm{He}_2$ molecules} \label{new_techniques}

The first of these methods is being developed at Brown University
by Maris and his colleagues\cite{Maris}. The idea of the method
comes from the fact that an application of a negative pressure to
$^4$He leads to an increase of the radius of the ion bubble which
explodes, and its size grows substantially when the negative
pressure has reached the critical value, $P_c$. Furthermore, Maris
\emph{et al.} argue that since the pressure around the quantized
vortex is reduced due to the Bernoulli effect, the ion bubbles
trapped on the vortex have the size larger than those in the bulk
of $^4$He and explode at the critical pressure of smaller
magnitude (hence the possibility of direct distinguishing the
bubbles trapped on vortices from those in the bulk of helium). In
these experiments the pressure variations were produced by an
acoustic, ultrasonic transducer generating either focused or
planar (as in the last of cited works) sound wave. By choosing a
suitable frequency and amplitude of the emitted sound, bubbles can
be make so large that they can be visualized by the conventional
optical methods (e.g. by a laser beam and a photomultiplier). The
other advantage of this method is that it uses neither emission of
the electron beam nor the electric field to accelerate the ions.
The charged particles are those that already present in the bulk
of $^4$He, either as a result of direct ionization by e.g. cosmic
rays, or through the more complicated process involving
ionization, production of UV photons, and ejection of electrons
from the walls into helium by the photoelectric effect.

The absence of the applied electric field means that electron
bubbles, unless they are trapped on quantized vortices, will
faithfully follow the normal flow. Indeed, the results of
visualization showed that in the weak, laminar counterflow most of
the bubbles moved along the streamlines of the normal fluid, but
alongside those about 10\% of the observed electron bubbles
followed snakelike paths; these could be assumed the electron
bubbles trapped by vortices and sliding along vortex filaments.

So far no new information on quantum turbulence has been obtained
using this technique, but its relative simplicity has a potential
to make it a tool competing with the PIV and particle tracking
techniques.

Another, completely new visualization technique is currently being
developed by McKinsey and his co-workers\cite{McKinsey} at Yale
University. This technique employs metastable $\textrm{He}_2$
triplet molecules, produced e.g. by the radioactive source, which
can be excited by two infrared photons from the ground
$a^3\Sigma_u^+$ state to the $d^3\Sigma_u^+$. About 90\% of the
excited molecules will decay into $b^3\Pi_g$ state emitting
photons at $640\,\textrm{nm}$ well separated in the wavelength
from the excitation photons and hence can be detected by standard
techniques. All excited molecules decay back to the ground state
within about $50\,\textrm{ns}$, so that the process can be
repeated many times to make possible a detection of a single
molecule.

The authors claim that $\textrm{He}_2$ molecules should be
unaffected by quantized vortices at temperatures above
$1\,\textrm{K}$ and so allow the resolution of the normal flow
even at the Kolmogorov scale. However, Vinen
commented\cite{Vinen-triplet_trapping} that at sufficiently low
temperatures triplet molecules may as well become trapped on
quantized vortex core. In this case three-dimensional images of
vortex lines can be produced by e.g. stereoscopic imaging.

If successful, the further development of this method may be
useful for recovering the velocity field and probability
distribution function of the normal velocity fluctuations, and can
also provide a new, nonintrusive tool for visualization of
quantized vortices.

\section{Conclusions} \label{conclusions}

This article reviewed the recent progress in understanding the motion of solid
particles and mechanisms of their interactions with both the normal fluid 
and quantized vortices
in turbulent $^4$He. This problem is addressed in the context of the PIV 
and the particle tracking techniques recently implemented in $^4$He, 
although we were certainly biased (perhaps, not surprisingly, 
considered our own research expertise)
towards theoretical and numerical works and interpretation of 
experimental results.

The problem
of the interaction of particle tracers with the normal fluid and with
quantized vortices is difficult, at least in principle:
it is three-dimensional, time dependent and strongly nonlinear. This is why
the best model available (the two-way coupling model described in Section 3) 
is computationally very expensive. In this review we have shown that,
fortunately, some aspects of this particle-vortex interaction can be 
understood using simpler models, either numerical or analytical, particularly 
given the insight gained from the two-way coupling model.

Although implemented recently (the first publications date 
2002), the PIV and the particle tracking techniques applied to the
problem of quantum turbulence in $^4$He had already produced some new
results. Among them: the discovery of large-scale eddies, which do not
have an analogue in the classical fluid dynamics, in the thermal
counterflow past a cylinder; the wider and flatter, compared to
classical, velocity profile in the turbulent boundary layer; scaling
for the time evolution of the parameter characterizing reconnections
of quantized vortices; non-Gaussian probability distribution function,
with the tail scaled as $v^{-3}$, of the velocity i decaying quantum
turbulence. These results seem to suggest that both the PIV and the
particle tracking techniques have great potential for studying quantum
turbulence. Besides, the motion of solid particles in turbulent $^4$He
seems to be itself an interesting and non-trivial phenomenon worth the
detailed theoretical and experimental study.

Although the review of this rapidly development area of research cannot be
comprehensive, we attempted to resolve, where possible, contradictions
in theoretical interpretations of recent
experimental observations and to address yet unresolved issues.

We also reviewed the related experimental methods, 
first of all the ion trapping technique which is perhaps the main tool
for studying quantized vortices in the low temperature limit. We argue
that these two experimental methods, i.e. the PIV/particle tracking and
the ion trapping techniques do not compete but complement each other.
Indeed, each of these techniques is most efficient in its own 
temperature and flow regime.

\begin{acknowledgements} We are grateful to our collaborators, W.F. Vinen and D.
Kivotides who contributed to some of the original research reviewed in this article,
and to K.R.~Sreenivasan for permission to use, in Sec.~\ref{mutual_friction}, the
results of yet unpublished experimental results of Bewley, Lathrop, Sreenivasan, and
Paoletti.

\end{acknowledgements}

%\pagebreak

\end{document}